\documentclass[oneside]{cernyrep}
\usepackage[T1]{fontenc}
\usepackage[bookmarks, colorlinks=true, linktoc=page, pdftex, linkcolor=black, citecolor=black, urlcolor=blue]{hyperref}

\usepackage{fancyhdr}
\fancyhfoffset{4 mm}
\fancypagestyle{ARTTITLE}{%
\fancyhf{} 
\lhead{\hfill CERN Accelerator School Proceedings ---
{\it Advanced Accelerator Physics} ---  Spa,  Belgium, 2024\hfill}
\lfoot{\hspace{3mm} Available online at \url{https://cas.web.cern.ch/previous-schools}}
\rfoot{\thepage\hspace*{3mm}}
 
}

\usepackage{varwidth}
\usepackage{xcolor}
\usepackage[labelfont=bf]{caption}

\usepackage{graphicx}
\usepackage{amssymb}
\usepackage{epstopdf}
\usepackage{listings}
\usepackage{color}
\usepackage[utf8]{inputenc}
\usepackage{pgfplots}
\usetikzlibrary{pgfplots.groupplots}
\usepackage{empheq}
\usepackage{subcaption}
\pgfplotsset{compat=1.18} 

\definecolor{codegreen}{rgb}{0,0.6,0}
\definecolor{codegray}{rgb}{0.5,0.5,0.5}
\definecolor{codepurple}{rgb}{0.58,0,0.82}
\definecolor{backcolour}{rgb}{0.95,0.95,0.92}
 
\lstdefinestyle{mystyle}{
    backgroundcolor=\color{backcolour},   
    commentstyle=\color{codegreen},
    keywordstyle=\color{magenta},
    numberstyle=\tiny\color{codegray},
    stringstyle=\color{codepurple},
    basicstyle=\footnotesize,
    breakatwhitespace=false,         
    breaklines=true,                 
    captionpos=b,                    
    keepspaces=true,                 
    numbers=left,                    
    numbersep=5pt,                  
    showspaces=false,                
    showstringspaces=false,
    showtabs=false,                  
    tabsize=2,
    basicstyle=\ttfamily\footnotesize
}

\usepackage{hyperref}
\usepackage{tikz}
\usepackage{tikz-3dplot}
\usetikzlibrary{shapes,arrows}
\newcommand{\blue}[1]{{\bf #1}}

\newcommand{\red}[1]{{\bf #1}}

\newcommand{\black}[1]{\textcolor{black}{#1}}

\tikzstyle{block} = [rectangle, draw, fill=blue!20, 
    text width=5em, text centered, rounded corners, minimum height=4em]
\tikzstyle{line} = [draw, -latex']

\begin{document}

\title{Introduction to Optics Design}
\author{G. Sterbini}
\institute{CERN, Geneva, Switzerland}

\begin{abstract}
This lecture provides an overview of the principles and methodologies involved in linear optics design. It aims to introduce key concepts such as the~matrix formalism, the symplecticity, the quantities that are preserved in the~single particle evolution, e.g.,~the Courant-Snyder invariant. It also covers the~concept of beam emittance and matching conditions for an ensemble of particles. The~goal is to equip readers with the foundational tools needed for both theoretical understanding and practical application in linear optics and accelerator design.
\end{abstract}
\keywords{Linear optics; symplecticity; twiss parameters; Courant-Snyder invariant; betatron oscillations; dispersion; chromaticity; sigma matrix; matched distribution; emittance.}
\maketitle
\thispagestyle{ARTTITLE}

\section{Introduction}

The linear optics theory was developed more than 65 years ago \cite{CS} and its first success was to demonstrate the overall focusing effect of a sequence of alternating focusing and defocusing quadrupoles (the so-called alternating-gradient principle).

Despite being based on simple linear algebraic concepts, the alternating-gradient principle was a~breakthrough in the history of accelerators. Since then, the \emph{linear optics design} of an accelerator is the very first step for its construction and for understanding the single particle motion. In addition, the linear optics serves as foundation to study the non-linear behaviour of a lattice, hence the importance to acquire a solid knowledge and familiarity with its concepts and the associated numerical methods. Presently, the~main challenges of the accelerator beam dynamics resides elsewhere (e.g., in the description of long-term behaviour of non-linear system). Therefore, this Chapter has to be intended as an introduction to the~problems presented in the following ones \cite{YP1, YP2}.
There is a rich bibliography covering the linear optics subject,  and hereby we indicated only a partial list of references \cite{CAS, LEE,  PEGGS, AW, CHAO}. 

The goal of the linear optics (and more in general of the beam dynamics) is to describe the motion of the~particles travelling in the accelerator or in a transfer line. The \emph{linear} adjective refers to the assumption, or the approximation, that the variation of the particle coordinates depends linearly on the coordinates themselves.

To introduce the linear optics theory, three  equivalent directions can be followed:
\begin{enumerate}
\item Integrating the equations of the motion. This is the historical approach and presents several limits when trying to generalize it to non-linear systems.
\item Using the Hamiltonian formalism to describe the particle motion. This approach is the natural one to  generalize the solutions to  non-linear dynamics problems \cite{YP1, YP2}.
\item Using a  computational approach,  deriving the linear optics theory using linear algebra concepts. This is the approach behind the standard  optics codes \cite{xsuite} and it is the one we will recall.
\end{enumerate}

\subsection{The reference system}
\label{sec:referenceSystem}
To describe the particle motion along the accelerator, it is needed to associate to each particle a set of coordinates with respect to a specific reference system and describe the evolution of this set of coordinate in time. 

Several reference frames can be chosen, e.g., a laboratory reference frame
\begin{equation}
[\mathcal{X}_0,\mathcal{Y}_0,\mathcal{Z}_0].
\end{equation}
The laboratory frame is not convenient to describe, in an efficient way, the particle motion. In fact, we can significantly simplify the problem by expressing the motion as relative to a so-called \emph{reference particle}. In other words, we can choose the reference frame,
\begin{equation}
[x_0,y_0,z_0],
\end{equation}
co-moving with the reference particle. 
In Figure~\ref{fig:referenceSystem}, the laboratory reference system $\{\mathcal{X}_0, \mathcal{Y}_0, \mathcal{Z}_0\}$, is shown in blue, and the co-moving reference system $\{x_0(s), y_0(s), z_0(s)\}$, in black.
It is worth noting that the latter depends on $s$, the longitudinal abscissa on the \emph{reference orbit} (black line). The reference orbit is described, by definition, by the origin of the $\{x_0(s), y_0(s), z_0(s)\}$ frame. In single-passage machines, like in LINACs or in transfer lines, we will call it \emph{reference trajectory}). The \emph{closed orbit} (in red, see Section~\ref{COComputation}) and \emph{betatron oscillations} (in orange, see Section~\ref{BetatronOscillation}) are then described with respect to the  $\{x_0(s), y_0(s), z_0(s)\}$ frame.

The main dipolar elements present in lattice (e.g., the arc main dipole) will define the reference orbit/trajectory (that is the geometry of the machine). There are other dipoles, usually referred to as correctors,  not contributing to the  machine geometry but defining the \emph{closed orbit}.

\begin{center}
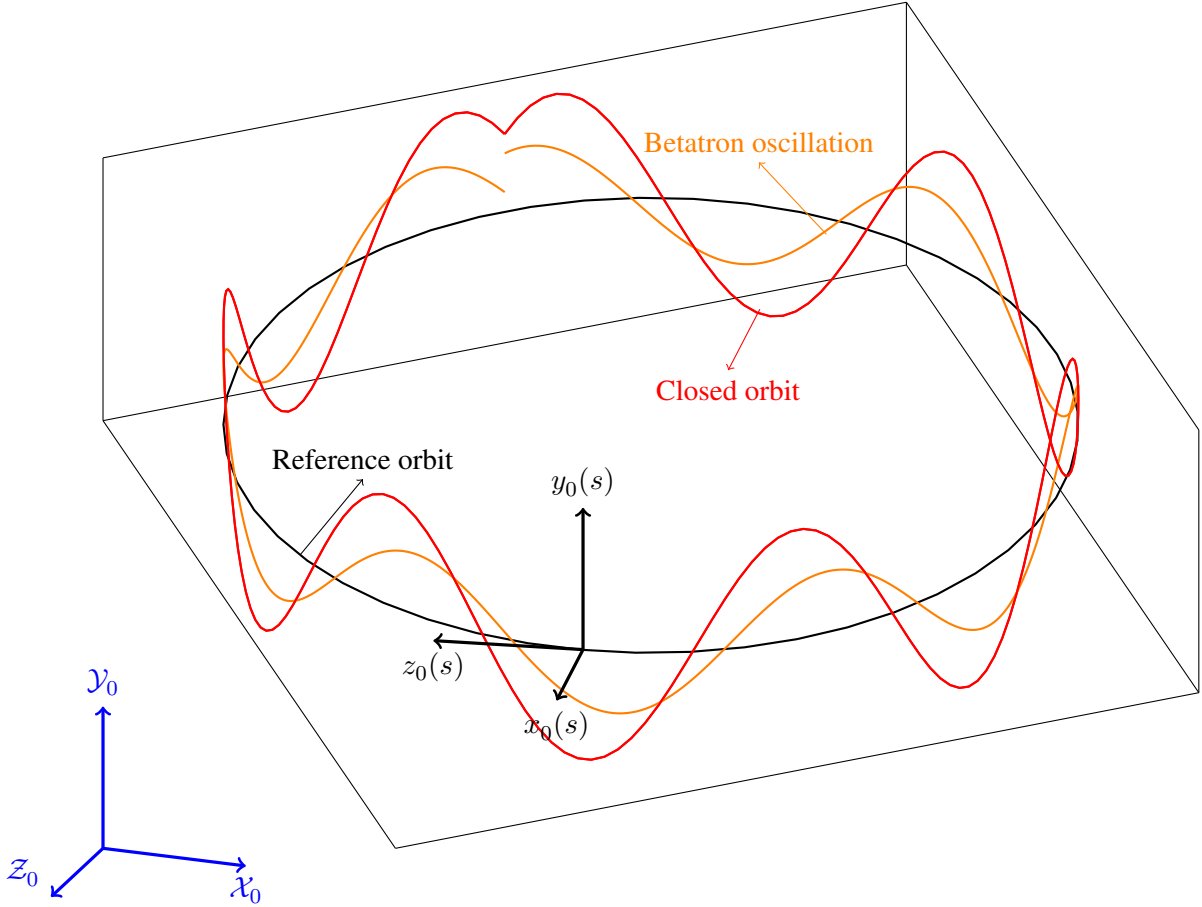
\begin{figure}
\tdplotsetmaincoords{70}{110}
\begin{tikzpicture}[scale=1, tdplot_main_coords]
  \begin{axis}[
   ticks=none,
   width=1\textwidth,
   height=0.8\textwidth,
   view = {70}{60}
  ]
  
  \addplot3 [
    samples=50,
    domain=0:360, 
    samples y=0,
    thick
  ] (
    {cos(x)},
    {sin(x)},
    {0}
  );
    \addplot3 [
    color=red,
    samples=200,
    domain=-180:180, 
    samples y=0,
    thick
  ] (
    {cos(mod(x,360)))},
    {sin(mod(x,360)))},
    {2*cos(6.25*mod(x,360))}
  );
  
  \addplot3 [
    color=orange,
    samples=600,
    domain=-180:180, 
    samples y=0,
    thick
  ] (
    {cos(mod(x,360)))},
    {sin(mod(x,360)))},
    {1*cos(6.25*mod(x,360))+.5*sin(6.25*x)}
  );
  
   \addplot3 [
    color=red,
    samples=200,
    domain=-180:180, 
    samples y=0,
    thick
  ] (
    {cos(mod(x,360)))},
    {sin(mod(x,360)))},
    {2*cos(6.25*mod(x,360))}
  );
\end{axis}

\tdplotsetrotatedcoords{0}{0}{90}
\draw[very thick,blue,tdplot_rotated_coords,->] (0,0,0) -- (2,0,0) node[anchor=north]{$\mathcal{X}_0$};
\draw[very thick,blue,tdplot_rotated_coords,->] (0,0,0) -- (0,-2,0) node[anchor=south east]{$\mathcal{Z}_0$};
\draw[very thick,blue,tdplot_rotated_coords,->] (0,0,0) -- (0,0,2) node[anchor=south]{$\mathcal{Y}_0$};

\tdplotsetrotatedcoords{0}{0}{10}
\coordinate (Shift) at (-9.4,3.3,0);
\tdplotsetrotatedcoordsorigin{(Shift)}

\draw[very thick,black,tdplot_rotated_coords,->] (0,0,0) -- (2,0,0) node[anchor=north ]{$x_0(s)$};
\draw[very thick,black,tdplot_rotated_coords,->] (0,0,0) -- (0,-2,0) node[anchor=north ]{$z_0(s)$};
\draw[very thick,black,tdplot_rotated_coords,->] (0,0,0) -- (0,0,2) node[anchor=south]{$y_0(s)$};

\draw[red,tdplot_rotated_coords,->] (-13.4,0,0) -- (-11,0,0) node[anchor=north ] {Closed orbit};
\draw[orange,tdplot_rotated_coords,->] (-15.6,0.5,.3) -- (-19,-1,0) node[anchor=south ] {Betatron oscillation};
\draw[black,tdplot_rotated_coords,->] (-3,-4.3,0) -- (-6,-4.,0) node[anchor=south ] {Reference orbit};

\end{tikzpicture}
\caption{Reference systems and reference orbit, closed orbit and betatron oscillations.
}
\label{fig:referenceSystem}
\end{figure}
\end{center}


\subsection{Linear symplectic transformations}

In the following, we denote with $X_s$
\begin{equation}
X_{s}=[x(s),p_x(s),y(s),p_y(s),z(s),p_z(s)]^T,
\end{equation}
the particle phase space coordinates at the $s$=position,
where we have a set of 6D canonical coordinates (e.g., the particle's positions and their corresponding momenta\footnote{In a dynamical system, we need positions and momenta to determine the system's evolution.}). With $X_{0,s}$, we refer to the closed orbit coordinates at the $s$-position.

We are interested in following the particle coordinates vector, $X$, from the position $s_1$ to the position $s_2$.
Our optics system is linear if and only if the evolution from the coordinates can be expressed as
\begin{equation}
\boxed{X_{s_2}-X_{0, s_2}=M\ (X_{s_1}-X_{0, s_1})}
\label{eq:linearity}
\end{equation}
where $M$ is a square matrix and does not depend on $X_{s_1}$ or $X_{s_2}$. Equation~\ref{eq:linearity} has to be valid for all $X$, $s_1$ and $s_2$\footnote{Very frequently, in Eq.~(\ref{eq:linearity}), the closed orbit is assumed to be on the reference orbit ($X_{0,s_1}=X_{0,s_2}=0$).}.

It is worth noting that, if $X_{s_1}=X_{0,~s_1}$, then $X_{s_2}=X_{0,~s_2}$ for all $M$. This means that, if the particle is on closed orbit at a given $s$, it stays on the closed orbit for all $s$.

In order to represent a physical transformation, in Eq.~(\ref{eq:linearity}), $M$ should be a \emph{symplectic matrix}. 
In the~following section, we will present the concept of symplecticity and its physical meaning. To do so, we need to introduce the concept of the bi-linear product. 

\subsection{The bi-linear product}

The \emph{bi-linear product} between the vectors $V$ and $U$ associated to the square matrix $F$ is defined as the~scalar  
\begin{equation}
V^T\ F\ U.
\end{equation}
As an example, the dot product is a bi-linear product with $F$ equal to the identity matrix.

It is interesting to study the properties of a linear transformation, $M$ (see Eq.~(\ref{eq:linearity})), that preserves the~bi-linear product associated with $F$. Observing that
 \begin{equation}
V^T\ F\ U= (M V)^T\ F\ M U  = V^T\ M^T\ F\ M\ U 
\end{equation}
we conclude that $M$ preserves the bi-linear product associated to $F$ if and only if
\begin{equation}
\boxed{F=M^T\ F\ M}.
\label{eq:preservation_of_bilinear_product}
\end{equation}

To be noted then, that if $M$ and $N$ are preserving the bi-linear product associated to $F$, then also $M\times N$ and $N\times M$ preserve it, therefore we can associate to a bi-linear product a group under matrix multiplication. In the following we will present two examples, the group of orthogonal and symplectic matrices.

\subsubsection{EXAMPLE. The orthogonal matrix}
Let us consider, for simplicity, the 2D case, that is,
 $U= (u_a, u_b)^T$ and $V= (v_a, v_b)^T$.
Assuming 
\begin{equation}
F=I=\left(
\begin{tabular}{cc}
    $1$ & $0$ \\
    $0$ & $1$\\
\end{tabular}
\right),
\end{equation}
the bi-linear transformation ${I}$ is the dot product between $V=(v_a, v_b)^T$  and $U=(u_a, u_b)^T$:
\begin{equation}
V^T\ \underbrace{{I}}_{F}\ U= v_a u_a+ v_b u_b .
\end{equation}
A matrix $M$ preserves the  bi-linear transformation  $I$  if and only if (see Eq.~(\ref{eq:preservation_of_bilinear_product}))
\begin{equation}
 M^T\ I\ M=I,
\end{equation}
then $M$ is called \emph{orthogonal} matrix. The physical meaning of an orthogonal matrix lies in the fact that it preserves the distance between vectors.

\subsubsection{EXAMPLE. The symplectic matrix}
Assuming
\begin{equation}
F={\Omega}=\left(
\begin{tabular}{cc}
    $0$ & $1$ \\
    $-1$ & $0$\\
\end{tabular}
\right),
\end{equation}
the bi-linear transformation  ${\Omega}$ is proportional to the amplitude of the cross product between $V=(v_a, v_b)^T$  and $U=(u_a, u_b)^T$:
\begin{equation}
V^T\ \underbrace{{\Omega}}_{F}\ U= v_a u_b-v_b u_a.
\end{equation}
The magnitude of the cross product can be interpreted as the positive area of the parallelogram having $U$ and $V$ as sides.
A matrix $M$ preserves the  bi-linear transformation  $\Omega$  (related to the cross product)  if and only if
\begin{equation}
\boxed{M^T\ \Omega\ M=\Omega},
\end{equation}
then $M$ is called \emph{symplectic} matrix.  The physical meaning of the symplectic matrix lies on the fact that it preserves the area between vectors.

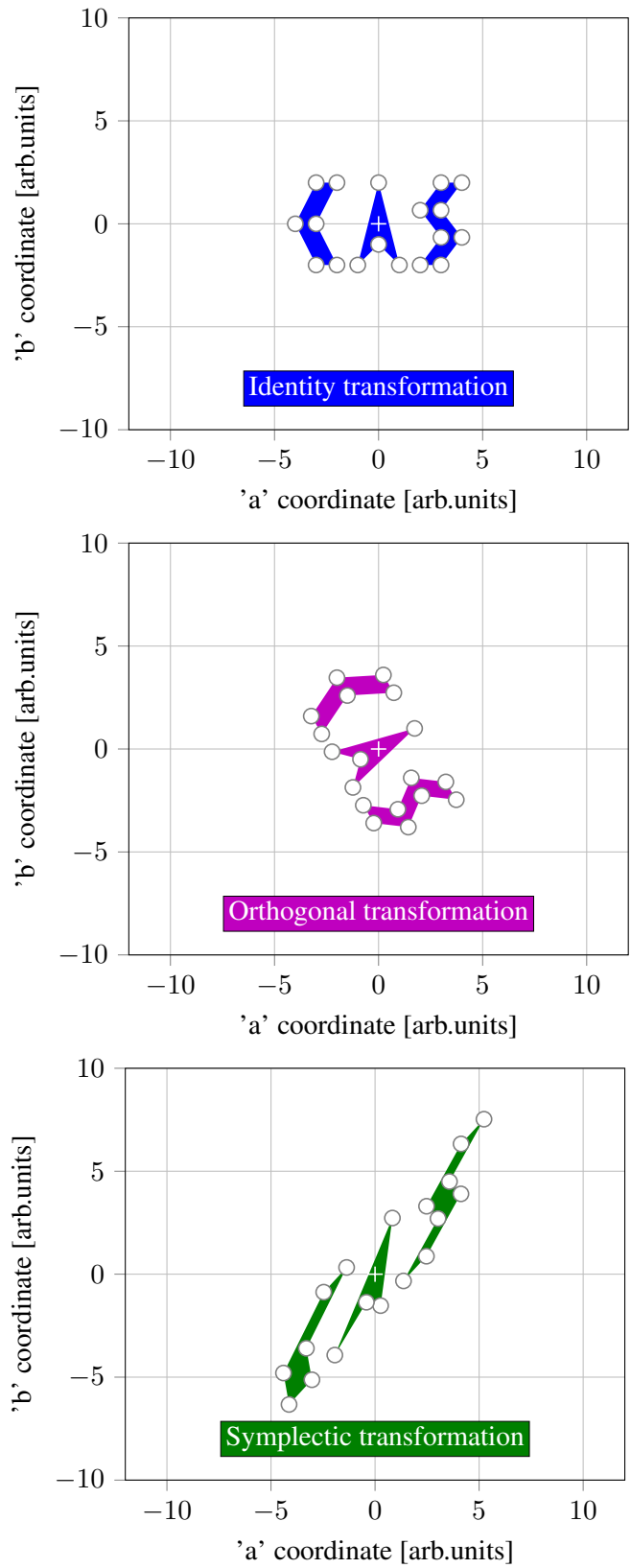
\begin{figure}
\begin{center}
\begin{tikzpicture}

\begin{axis}[
tick align=outside,
tick pos=left,
xlabel={'a' coordinate [arb.units]},
xmajorgrids,
xmin=-12, xmax=12,
ylabel={'b' coordinate [arb.units]},
ymajorgrids,
ymin=-10, ymax=10
]
\path [fill=blue, draw opacity=0] (axis cs:-3,-2)
--(axis cs:-2,-2)
--(axis cs:-3,0)
--(axis cs:-2,2)
--(axis cs:-3,2)
--(axis cs:-4,0)
--cycle;

\path [fill=blue, draw opacity=0] (axis cs:-1,-2)
--(axis cs:0,2)
--(axis cs:1,-2)
--(axis cs:0,-1)
--cycle;

\path [fill=blue, draw opacity=0] (axis cs:2,-2)
--(axis cs:3,-2)
--(axis cs:4,-0.66)
--(axis cs:3,0.66)
--(axis cs:4,2)
--(axis cs:3,2)
--(axis cs:2,0.66)
--(axis cs:3,-0.66)
--cycle;

\addplot [semithick, white, mark=*, mark size=3, mark options={solid,draw=white!50.19607843137255!black}, only marks, forget plot]
table [row sep=\\]{%
-3	-2 \\
-2	-2 \\
-3	0 \\
-2	2 \\
-3	2 \\
-4	0 \\
};
\addplot [semithick, white, mark=*, mark size=3, mark options={solid,draw=white!50.19607843137255!black}, only marks, forget plot]
table [row sep=\\]{%
-1	-2 \\
0	2 \\
1	-2 \\
0	-1 \\
};
\addplot [semithick, white, mark=+, mark size=3, mark options={solid}, only marks, forget plot]
table [row sep=\\]{%
0	0 \\
};
\addplot [semithick, white, mark=*, mark size=3, mark options={solid,draw=white!50.19607843137255!black}, only marks, forget plot]
table [row sep=\\]{%
2	-2 \\
3	-2 \\
4	-0.66 \\
3	0.66 \\
4	2 \\
3	2 \\
2	0.66 \\
3	-0.66 \\
};
\node at (axis cs:0,-8)[
  fill=blue,
  draw=black,
  line width=0.2pt,
  inner sep=2pt,
  text=white,
  rotate=0.0
]{ Identity transformation};
\end{axis}

\end{tikzpicture}
\begin{tikzpicture}

\definecolor{color0}{rgb}{0.75,0,0.75}

\begin{axis}[
tick align=outside,
tick pos=left,
xlabel={'a' coordinate [arb.units]},
xmajorgrids,
xmin=-12, xmax=12,
ylabel={'b' coordinate [arb.units]},
ymajorgrids,
ymin=-10, ymax=10
]
\path [fill=color0, draw opacity=0] (axis cs:-3.23205080756888,1.59807621135332)
--(axis cs:-2.73205080756888,0.732050807568877)
--(axis cs:-1.5,2.59807621135332)
--(axis cs:0.732050807568877,2.73205080756888)
--(axis cs:0.232050807568877,3.59807621135332)
--(axis cs:-2,3.46410161513775)
--cycle;

\path [fill=color0, draw opacity=0] (axis cs:-2.23205080756888,-0.133974596215561)
--(axis cs:1.73205080756888,1)
--(axis cs:-1.23205080756888,-1.86602540378444)
--(axis cs:-0.866025403784439,-0.5)
--cycle;

\path [fill=color0, draw opacity=0] (axis cs:-0.732050807568877,-2.73205080756888)
--(axis cs:-0.232050807568877,-3.59807621135332)
--(axis cs:1.42842323350227,-3.79410161513775)
--(axis cs:2.07157676649773,-2.26807621135332)
--(axis cs:3.73205080756888,-2.46410161513775)
--(axis cs:3.23205080756888,-1.59807621135332)
--(axis cs:1.57157676649773,-1.40205080756888)
--(axis cs:0.928423233502271,-2.92807621135332)
--cycle;

\addplot [semithick, white, mark=*, mark size=3, mark options={solid,draw=white!50.19607843137255!black}, only marks, forget plot]
table [row sep=\\]{%
-3.23205080756888	1.59807621135332 \\
-2.73205080756888	0.732050807568877 \\
-1.5	2.59807621135332 \\
0.732050807568877	2.73205080756888 \\
0.232050807568877	3.59807621135332 \\
-2	3.46410161513775 \\
};
\addplot [semithick, white, mark=*, mark size=3, mark options={solid,draw=white!50.19607843137255!black}, only marks, forget plot]
table [row sep=\\]{%
-2.23205080756888	-0.133974596215561 \\
1.73205080756888	1 \\
-1.23205080756888	-1.86602540378444 \\
-0.866025403784439	-0.5 \\
};
\addplot [semithick, white, mark=+, mark size=3, mark options={solid}, only marks, forget plot]
table [row sep=\\]{%
0	0 \\
};
\addplot [semithick, white, mark=*, mark size=3, mark options={solid,draw=white!50.19607843137255!black}, only marks, forget plot]
table [row sep=\\]{%
-0.732050807568877	-2.73205080756888 \\
-0.232050807568877	-3.59807621135332 \\
1.42842323350227	-3.79410161513775 \\
2.07157676649773	-2.26807621135332 \\
3.73205080756888	-2.46410161513775 \\
3.23205080756888	-1.59807621135332 \\
1.57157676649773	-1.40205080756888 \\
0.928423233502271	-2.92807621135332 \\
};
\node at (axis cs:0,-8)[
  fill=color0,
  draw=black,
  line width=0.2pt,
  inner sep=2pt,
  text=white,
  rotate=0.0
]{ Orthogonal transformation};
\end{axis}

\end{tikzpicture}
\begin{tikzpicture}

\begin{axis}[
tick align=outside,
tick pos=left,
xlabel={'a' coordinate [arb.units]},
xmajorgrids,
xmin=-12, xmax=12,
ylabel={'b' coordinate [arb.units]},
ymajorgrids,
ymin=-10, ymax=10
]
\path [fill=green!50.0!black, draw opacity=0] (axis cs:-4.13333333333333,-6.32727272727273)
--(axis cs:-3.03333333333333,-5.12727272727273)
--(axis cs:-3.3,-3.6)
--(axis cs:-1.36666666666667,0.327272727272727)
--(axis cs:-2.46666666666667,-0.872727272727273)
--(axis cs:-4.4,-4.8)
--cycle;

\path [fill=green!50.0!black, draw opacity=0] (axis cs:-1.93333333333333,-3.92727272727273)
--(axis cs:0.833333333333333,2.72727272727273)
--(axis cs:0.266666666666667,-1.52727272727273)
--(axis cs:-0.416666666666667,-1.36363636363636)
--cycle;

\path [fill=green!50.0!black, draw opacity=0] (axis cs:1.36666666666667,-0.327272727272727)
--(axis cs:2.46666666666667,0.872727272727273)
--(axis cs:4.125,3.9)
--(axis cs:3.575,4.5)
--(axis cs:5.23333333333333,7.52727272727273)
--(axis cs:4.13333333333333,6.32727272727273)
--(axis cs:2.475,3.3)
--(axis cs:3.025,2.7)
--cycle;

\addplot [semithick, white, mark=*, mark size=3, mark options={solid,draw=white!50.19607843137255!black}, only marks, forget plot]
table [row sep=\\]{%
-4.13333333333333	-6.32727272727273 \\
-3.03333333333333	-5.12727272727273 \\
-3.3	-3.6 \\
-1.36666666666667	0.327272727272727 \\
-2.46666666666667	-0.872727272727273 \\
-4.4	-4.8 \\
};
\addplot [semithick, white, mark=*, mark size=3, mark options={solid,draw=white!50.19607843137255!black}, only marks, forget plot]
table [row sep=\\]{%
-1.93333333333333	-3.92727272727273 \\
0.833333333333333	2.72727272727273 \\
0.266666666666667	-1.52727272727273 \\
-0.416666666666667	-1.36363636363636 \\
};
\addplot [semithick, white, mark=+, mark size=3, mark options={solid}, only marks, forget plot]
table [row sep=\\]{%
0	0 \\
};
\addplot [semithick, white, mark=*, mark size=3, mark options={solid,draw=white!50.19607843137255!black}, only marks, forget plot]
table [row sep=\\]{%
1.36666666666667	-0.327272727272727 \\
2.46666666666667	0.872727272727273 \\
4.125	3.9 \\
3.575	4.5 \\
5.23333333333333	7.52727272727273 \\
4.13333333333333	6.32727272727273 \\
2.475	3.3 \\
3.025	2.7 \\
};
\node at (axis cs:0,-8)[
  fill=green!50.0!black,
  draw=black,
  line width=0.2pt,
  inner sep=2pt,
  text=white,
  rotate=0.0
]{ Symplectic transformation};
\end{axis}

\end{tikzpicture}
\caption{Examples of orthogonal and symplectic transformations.}
\label{fig:example1}
\end{center}
\end{figure}

In Figure~\ref{fig:example1} we can see a graphical representation of an orthogonal and a symplectic linear transformation. Comparing two generic vectors between the identity transformation (upper plot) and the orthogonal transformation (middle plot) one can note that their dot product is conserved (i.e.,~\emph{all distances} are preserved). In the symplectic transformation (lower plot) given a  set on $n>3$ vectors (defining a polygon) the surface of the polygon is preserved. For that reason a symplectic linear transformation preserves the~phase-space areas. Behind this concept lies a much more general theorem of classical mechanics valid also in the non-linear case, the \emph{Liouville theorem}.

\subsubsection{Properties of the symplectic matrices}
All the concepts we introduced so far  can be generalized from the 2D to 4D and 6D. In particular, in 6D, $\Omega$ becomes a $6\times6$ matrix\footnote{In some texts the symplectic matrix is expresses as \[
\Omega =
\begin{bmatrix}
0 & I_n \\
- I_n & 0
\end{bmatrix}.
\]}:

\begin{equation}
\Omega=
\begin{pmatrix}
0 & 1 & 0 & 0 & 0 & 0 \\
-1 & 0 & 0 & 0 & 0 & 0 \\
0 & 0 & 0 & 1 & 0 & 0 \\
0 & 0 & -1 & 0 & 0 & 0 \\
0 & 0 & 0 & 0 & 0 & 1 \\
0 & 0 & 0 & 0 & -1 & 0
\end{pmatrix}.
\end{equation}
An example of symplectic matrix in 2D (not block-symplectic) is the following
\begin{equation}
\begin{pmatrix}
1 & 0 & 0 & 0\\
0 & 1 & 1 & 0\\
0 & 0 & 1 & 0\\
1 & 0 & 0 & 1
\end{pmatrix}.
\end{equation}

It is worth recalling some important properties of the symplectic group:
\begin{itemize}
\item as already mentioned, if $M_1$ and  $M_2$ are symplectic then $M=M_1 M_2$ is symplectic too. 
\item If $M$ is symplectic then $M^T$ is symplectic. 
\item Every symplectic matrix is invertible
\begin{equation}
M^{-1}=\Omega^{-1} M^{T} \Omega
\label{eq:inversion}
\end{equation}
and  $M^{-1}$ is symplectic. Therefore an inversion of a symplectic matrix can be very efficient in term of computational cost. 
\item A necessary (but not sufficient) condition for M to be symplectic is that $\det(M)=+1$. For the~2D case, this condition is necessary and sufficient. An example of not symplectic matrix M with $\det(M)=+1$ is the following
\begin{equation}
\begin{pmatrix}
1 & 0 & 1 & 0\\
0 & 1 & 1 & 0\\
0 & 0 & 1 & 0\\
1 & 0 & 0 & 1
\end{pmatrix}.
\end{equation}
\item There are symplectic matrices that are defective, that is they cannot be diagonalized, e.g., 
\begin{equation}
\begin{pmatrix} 1 & 1 \\ 0 & 1\end{pmatrix}.
\end{equation}
\end{itemize}

\subsection{Symplectic matrix and accelerators}

From the concept of symplectic transformation we can define the basic building blocks that constitute all the linear transformation in an accelerator.
In particular for the 2D case, we can consider the following three matrices:
\begin{equation}
\underbrace{\begin{pmatrix}
   G & 0 \\[6pt]
    0 & \dfrac{1}{G} \\
\end{pmatrix}}_{\text{thin telescope}},\   
\underbrace{
\begin{pmatrix}
    1 & L \\[12pt]
    0 & 1\\
\end{pmatrix}}_{\text{{drift}}},\ 
\underbrace{
\begin{pmatrix}
    1 & 0 \\[6pt]
    -\dfrac{1}{f} & 1\\
\end{pmatrix}
}_{\text{{thin quad}}}.
\end{equation}

Among the above matrices, one can recognise the  $L$-long drift and the thin focusing quadrupole with focal length $f$. 
In addition, there is also the thin telescope matrix. This matrix reduces to the identical transformation for $G=1$, but for $G\ne1$ introduces a discontinuity in the position coordinate, while the~other thin matrix (the thin quadrupole) introduces a discontinuity only in the momentum coordinate (thin kick).
Conveniently combining these drifts and thin quadrupoles, one can find back also the well-known matrices for the 2D thick elements.

\subsubsection{EXAMPLE. A thick quadrupole}

One can derive the transfer matrix of a thick quadrupole of length $L$ and normalized gradient $K1$ by factorizing it in $n$ identical optics cells. Each cell is constituted by  a $\dfrac{L}{n}$-long drift and a thin quadrupole with focal length $\dfrac{n}{\text{K1}~L}$.  From this factorization, one can obtain the thick lens quadrupole matrix by  solving the following limit

\begin{eqnarray}
\lim_n \left[
\left(
\begin{tabular}{cc}
    1 & 0 \\[6pt]
    $-\dfrac{\text{K1}\ L}{n}$ & 1\\
\end{tabular}
\right)
\left(
\begin{tabular}{cc}
    1 & $\dfrac{L}{n}$  \\[6pt]
    0 & 1\\
\end{tabular}
\right)
\right]^n=\\
\left(
\begin{array}{cc}
 \cos \left(\sqrt{\text{K1}} L\right) & \dfrac{\sin \left(\sqrt{\text{K1}} L\right)}{\sqrt{\text{K1}}} \\[12pt]
 -\sqrt{\text{K1}} \sin \left(\sqrt{\text{K1}} L\right) & \cos \left(\sqrt{\text{K1}} L\right) \\
 \end{array}
\right).
\end{eqnarray}

To compute the above this limit and, in general, for symbolic computations one can profit of the available symbolic computation tools (e.g., Mathematica\texttrademark~or the Python package \textit{sympy}). An example of symbolic calculation of the above limit is given in Listing~\ref{lst:mathematicaLimit}.


We established a correspondence between elements along our machine (drift, bending, quadrupoles, solenoids,\dots) and symplectic matrices. For a rich list of matrix transformations in an accelerator refer to the Appendix in \cite{PEGGS}.

\subsection{Tracking in a linear system}
Given a sequence of elements $M_1, M_2$, \dots $M_k$ (the \emph{lattice}), the evolution of the coordinate, $X_n$, along the lattice for a given particle can be obtained as
\begin{equation}
X_n=M_n\dots M_1\ X_0\ {\rm for}\ n\ge1.
\label{eq:tracking}
\end{equation}

The transport of the particles along the lattice is called \emph{tracking}.  The tracking on a linear system is trivial and, as we will show in the following, unnecessary. In fact, we can  decompose the trajectory of the~single particle in terms of its invariant of the motion and of the properties of the lattice. Thanks to these quantities, we can describe not only the trajectory of the single particle but also the statistical evolution of  an ensemble of particles (the beam envelope).
So, instead of tracking an ensemble of particles, we will focus to define and compute the properties of the lattice, on one hand, and of the particle/beam, on the other.

\section{Linear Lattices}

\subsection{Periodic lattice and stability}
We study in this section the motion of the particles in a periodic lattice, that is a lattice constituted by an indefinite repetition of the same basic $C$-long period. We represent it with $M_{OTM}$, the so-called One-Turn-Map, that is the linear symplectic matrix of a single turn. Due to its periodicity we have:
\begin{equation}
M_{OTM}(s_0)=M_{OTM}(s_0+C).
\end{equation}
From Eq.~(\ref{eq:tracking}) we get
\begin{equation}
X_m=M_{OTM}^m\ X_0,
\end{equation}
where we used the subscript $m$ to refer to the turn number. In the following we study the property of $M_{OTM}$ to have stable motion in the lattice. The stable motion implies that the motion of a generic initial condition, $X_0$, stays bounded,  that is we can always define a finite $\hat{X}$ such that 
\begin{equation}
|X_m|< |\hat{X}| \ {\rm for\ all}\ X_0\ {\rm and}\  m.
\end{equation}

In other words, to verify if the lattice is stable we need to verify that all the elements of the matrix $M_{OTM}^m$ stay bounded for $m\rightarrow\infty$ . To solve this problem we use in the following three equivalent factorization forms:
\begin{itemize}
\item diagonal-factorization,
\item R-factorization,
\item Twiss-factorization.
\end{itemize}

\subsubsection{Diagonal-factorization}

If  $M_{OTM}$ can be expressed as a diagonal-factorization (e.g., in diagonal form) 
\begin{equation}
M_{OTM}=P 
\underbrace{
\begin{pmatrix}
  \lambda_1 &         0  \\
          0 & \lambda_2 
\end{pmatrix}}_\text{D} 
 P^{-1},
\end{equation}
after $m$-turns, we have that
\begin{equation}
M_{OTM}^m=
\underbrace{P D P^{-1}}_1\times
\underbrace{P D P^{-1}}_2\times
\dots\times
\underbrace{P D P^{-1}}_m=P D^m P^{-1}.
\end{equation}
Therefore, the stability depends only on the eigenvalues of $M_{OTM}$.

Note that  if $V$ is an eigenvector also $k 
\times V,\ k\ne0$  is an eigenvector. Therefore \emph{P is not uniquely defined}: we chose $P$ such that $\det(P)=-i$. There is not an physical meaning behind this convention, but as it will appear clearly later, it is convenient since is compatible with standard definitions of the~accelerator dynamics.
It is worth recalling the following properties:
\begin{itemize}
\item for a real matrix the eigenvalues, if complex, appear in complex conjugate pairs. 
\item For a symplectic matrix  $M_{OTM}$
\begin{equation}
\prod_i^{2n} \lambda_i=1,
\end{equation}
where $\lambda_i$ are the eigenvalues of $M_{OTM}$. 
\item Therefore, for 2$\times$2 symplectic matrix, the eigenvalues can be written as $\lambda_1=e^{i\ \mu_{OTM}}$ and $\lambda_2=e^{-i\ \mu_{OTM}}$ (without loss of generality we consider $\mu_{OTM}>0$). This implies that 
\begin{equation}
\boxed{D^m=D(m\mu_{OTM})}.
\end{equation}Therefore, a power of a matrix is reduced to a simple scalar multiplication.
\end{itemize}

\emph{If $\mu$ is real then the motion is stable} and we can define the \emph{fractional tune} of the periodic lattice as $\dfrac{\mu_{OTM}}{2\pi}$. We will describe in Section~\ref{sec:betatronPhase} how to compute the total phase advance of the machine and, therefore, the \emph{integer tune}. 

\subsubsection{R-factorization}

The diagonal-factorization is  convenient to check the stability but not to visualize the turn-by-turn phase-space evolution of the particle.
To do that it is worth considering the rotation-factorization
\begin{equation}
\boxed{M_{OTM} = \bar{P}  
\underbrace{
\begin{pmatrix}
  \cos\mu_{OTM} &          \sin\mu_{OTM}  \\[6pt]
  -\sin\mu_{OTM} &  \cos\mu_{OTM}  
\end{pmatrix}}_\text{R($\mu_{OTM}$) is orthogonal} 
 \bar{P}^{-1}}.
\label{eq:rfactorization}
\end{equation}
If $M_{OTM}$ can be diagonalized then can be expressed also in a R-factorization. In fact, to go from diagonal-factorization to R-factorization we note that 

\begin{equation}
\underbrace{\begin{pmatrix}
  \cos\mu_{OTM} &          \sin\mu_{OTM}  \\[6pt]
  -\sin\mu_{OTM} &  \cos\mu_{OTM}  
\end{pmatrix}}_{R(\mu_{OTM})}=
\underbrace{
\begin{pmatrix}
  \dfrac{1}{\sqrt{2}} &          \dfrac{1}{\sqrt{2}}  \\[12pt]
  \dfrac{i}{\sqrt{2}} &  -\dfrac{i}{\sqrt{2}}
\end{pmatrix}}_{S^{-1}}
\underbrace{
\begin{pmatrix}
 e^{i\ \mu_{OTM}} &        0\\[6pt]
  0 &  e^{-i\ \mu_{OTM}} 
\end{pmatrix}
}_{D(\mu_{OTM})}
\underbrace{
\begin{pmatrix}
  \dfrac{1}{\sqrt{2}} &          -\dfrac{i}{\sqrt{2}}  \\[12pt]
  \dfrac{1}{\sqrt{2}} &  \dfrac{i}{\sqrt{2}}
\end{pmatrix}}_{S},
\end{equation}
where we introduced the matrix $S$. Therefore,

\begin{equation}
R^m=R(m\mu_{OTM}).
\end{equation}
One can easily express $\bar{P}$ as function of $P$ and $S$ observing that 
\begin{equation} 
M_{OTM} = 
\underbrace{P\ 
S}_{\bar{P}}
\ \underbrace{S^{-1}\ D\
S}_R\ \underbrace{S^{-1}\ P^{-1}}_{\bar{P}^{-1}},
\end{equation}
i.e., $\bar{P}=P S$.
We note that by choosing $\det(P)=-i$ we got $\det(\bar{P})=1$, that is we expressed \emph{$M$ as the product of orthogonal and symplectic matrices}.  This result is very relevant, since it implies that the~\emph{$M_{OTM}$ is similar to a pure rotation}. That is, with a convenient change of base (expressed by the~matrix $\bar{P}^{-1}$), we can move from the physical phase-space to the \emph{normalized phase-space} where the~periodic motion is just a clockwise rotation of the angle $\mu_{OTM}$.

\subsubsection{Twiss-factorization of  $M_{OTM}$}
We note that 
\begin{equation}
R(\mu_{OTM})=
\begin{pmatrix}
 1 &        0\\
  0 &  1
\end{pmatrix}
\cos\mu_{OTM} +
\begin{pmatrix}
 0 &        1\\
 -1 &  0
\end{pmatrix}
\sin{\mu_{OTM}},
\end{equation}
yielding the, so called, Twiss-factorization 
\begin{equation}
\boxed{M_{OTM}=\underbrace{\bar{P} I  \bar{P}^{-1}}_I\cos\mu_{OTM} + \underbrace{\bar{P} \Omega  \bar{P}^{-1}}_{\blue{J}} \sin\mu_{OTM}}.
\end{equation}
It is worth observing that $J$ has three  properties: 
\begin{enumerate}
\item $\det(J)=1$,
\item $J_{11}=-J_{22}$,
\item  $J_{12}>0$.
\end{enumerate}
The last two expressions can be proven using the symbolic computation as show in Listing~\ref{lst: J}.



The following parametric expression of $J$ has been proposed \cite{CS}
\begin{equation}
\boxed{J=\begin{pmatrix}
 \blue{\alpha} &        \overbrace{\blue{\beta}}^{\red{>0}}\\[6pt]
-\underbrace{\dfrac{1+\blue{\alpha}^2}{\blue{\beta}}}_{\blue{\gamma}\red{>0}} &  -\blue{\alpha}
\end{pmatrix}}
\end{equation}
defining the \emph{Twiss parameters}, $\alpha,\ \beta,\ \gamma$ of the lattice at the start of the sequence $M_{OTM}$.
It is  important to note that they are not depending on the turn number $m$ since
\begin{equation}
M_{OTM}^m= I \cos(m\mu_{OTM})+J\sin(m\mu_{OTM}).
\end{equation}

In other words the \emph{Twiss parameters in a stable periodic lattice are periodic}.
From the definition of $J$, $J=\bar{P} \Omega \bar{P}^{-1}$,  we can express $\bar{P}$, $\bar{P}^{-1}$ and $P$ as function of the Twiss parameters:
\begin{equation}
\bar{P}=\red{\begin{pmatrix}
 \sqrt{\beta} &        0\\[6pt]
-\dfrac{\alpha}{\sqrt{\beta}} &  \dfrac{1}{\sqrt{\beta}}
\end{pmatrix}}=
\begin{pmatrix}
 \sqrt{\beta} &        0\\[6pt]
0 &  \dfrac{1}{\sqrt{\beta}}
\end{pmatrix}
\begin{pmatrix}
 1 &        0\\[6pt]
-\dfrac{\alpha}{\sqrt{\beta}} &  1
\end{pmatrix},
\end{equation}

\begin{equation}
\bar{P}^{-1}=
\begin{pmatrix}
 \dfrac{1}{\sqrt{\beta}} &        0\\[12pt]
\dfrac{\alpha}{\sqrt{\beta}} &  \sqrt{\beta}
\end{pmatrix},
\end{equation}
and
\begin{equation}
P=\bar{P} S^{-1}=\red{\left(
\begin{array}{cc}
\sqrt{\dfrac{\beta }{2}} &  \sqrt{\dfrac{\beta }{2}}\\[12pt]
\dfrac{-\alpha +i}{\sqrt{2\beta }} &   \dfrac{-\alpha -i}{\sqrt{2\beta}}\\
\end{array}
\right).}
\label{eq:p}
\end{equation}

To summarise,  if the matrix $M_{OTM}$ is diagonalizable and if $|\lambda_1|$=1, the lattice is stable and its fractional tune is  $\dfrac{|{\rm phase}( \lambda_1)|}{2\pi}$. From the eigenvector matrix $P$, conveniently normalized with $\det{P}=-i$, and from Eq.~(\ref{eq:p}), one can find the Twiss parameters of the lattice at the $M_{OTM}$ starting point. In Listing~\ref{lst: basicLinearOptics}, an example for computing the optical functions at the $M_{OTM}$ starting point is shown.


\subsection{Twiss parameters along the machine}

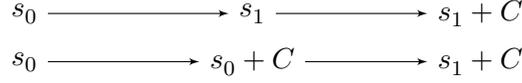
\begin{figure}
\begin{center}
\begin{tikzpicture}[node distance = 3cm, auto]
    \node (step1) {{$s_0$}};
    \node [right of= step1] (step2) {$\black{s_1}$};
    \node [right of= step2] (step3) {$\black{s_1+C}$};
    \path [line] (step1) -- (step2);
    \path [line] (step2) -- (step3);
\end{tikzpicture}
\\

\begin{tikzpicture}[node distance = 3cm, auto]
    \node (step1) {${s_0}$};
    \node [right of= step1] (step2) {$\black{s_0+C}$};
    \node [right of= step2] (step3) {$\black{s_1+C}$};
    \path [line] (step1) -- (step2);
    \path [line] (step2) -- (step3);
\end{tikzpicture}
\end{center}
\caption{Similarity of two One-Turn-Map matrices referred to two different points $s_0$ and $s_1$.}
\label{fig:similarity}
\end{figure}

Given a C-long periodic lattice and two longitudinal positions $s_0$ and $s_1$ ($s_1>s_0$), as depicted in Fig.~\ref{fig:similarity}, the transformation from $s_0$ to $s_1+C$ can be expressed as

\begin{equation}
\black{M_{OTM}(s_1)\ M} =\black{M\ M_{OTM}(s_0)}
\end{equation}
where $M$ is the transport matrix from $s_0$ to $s_1$. This implies 
\begin{equation}
\boxed{{M_{OTM}(s_1)=M\ M_{OTM}(s_0)\ M^{-1}}},
\end{equation}
that is the matrices $M_{OTM}(s_1)$ and $M_{OTM}(s_2)$ are similar and therefore they have the same eigenvalues. From this observation it yields that the $M_{OTM}$ is $s$-dependent but the tune is not.

\subsubsection{$\beta$  and $\alpha$ transport}
We  observe that $\beta$ and $\alpha$ are s-dependent functions. In fact we have:
\begin{equation}
M_{OTM}(s_1)=M \ M_{OTM}(s_0)\ M ^{-1}=M \ (I \cos\mu_{OTM}  + J(s_0) \sin\mu_{OTM})\ M ^{-1},
\end{equation}
therefore
\begin{equation}
\underbrace{\begin{pmatrix}
\alpha(s_1) &        \beta(s_1)\\[6pt]
-\gamma(s_1) &  -\alpha(s_1)
\end{pmatrix}}_{J(s_1)}
 = M  
\underbrace{
\begin{pmatrix}
\alpha(s_0) &        \beta(s_0)\\[6pt]
-\gamma(s_0)&  -\alpha(s_0)
\end{pmatrix}}_{J(s_0)} \ M^{-1}.
\label{eq:transport}
\end{equation}
From  Eq.~(\ref{eq:inversion})  (inverse of a symplectic matrix) we have
\begin{equation}
\begin{pmatrix}
\alpha(s_1) &        \beta(s_1) \\[6pt]
-\gamma(s_1)  &  -\alpha(s_1) 
\end{pmatrix}
\Omega^{-1}
 = M  
\begin{pmatrix}
\alpha(s_0) &        \beta(s_0)\\[6pt]
-\gamma(s_0) &  -\alpha(s_0)
\end{pmatrix}\ \Omega^{-1}\ M^T,
\end{equation}
that is 
\begin{equation}
\boxed{\underbrace{{\begin{pmatrix}
\beta(s_1)  &        -\alpha(s_1) \\[6pt]
-\alpha(s_1)  & \gamma(s_1) 
\end{pmatrix}}}_{J(s_1)\ \Omega^{-1}}
 = M  
\underbrace{{\begin{pmatrix}
\beta(s_0)  &        -\alpha(s_0)\\[6pt]
-\alpha(s_0) & \gamma(s_0)
\end{pmatrix}}}_{J(s_0)\ \Omega^{-1}}
\ M^T}.
\label{eq:standardPropagation}
\end{equation}
Equation~(\ref{eq:standardPropagation}) allows us to propagate the initial condition of the optical functions $\beta$ and $\alpha$ along the~lattice. It is worth noting that from Eq.~(\ref{eq:standardPropagation}) and remembering the definition of the $\bar{P}^{-1}$ matrix (the one to transform the physical phase-space in the normalized phase-space), we get 
\begin{equation}
 \bar{P}^{-1}  
{{\begin{pmatrix}
\beta(s_0)  &        -\alpha(s_0)\\[6pt]
-\alpha(s_0) & \gamma(s_0)
\end{pmatrix}}}
\ \left(\bar{P}^{-1}\right)^T=
{{\begin{pmatrix}
1  &        0 \\[6pt]
0  & 1
\end{pmatrix}}},
\label{eq:normalizedPS}
\end{equation}hence, in the normalized phase-space, $\beta$ and $\alpha$ are 1 and 0, respectively.

\subsubsection{EXAMPLE. The $\beta$-function in a drift}
To compute the Twiss parameters in a drift starting from $\beta_0$  and $\alpha_0$, we can simply apply the Eq.~(\ref{eq:standardPropagation})
\begin{equation}
\begin{pmatrix}
\beta(s) &        -\alpha(s)\\[6pt]
-\alpha(s) & \gamma(s)
\end{pmatrix}= 
\begin{pmatrix}
1 &      s\\[6pt]
0 & 1
\end{pmatrix}
\begin{pmatrix}
\beta_0 &        -\alpha_0\\[6pt]
-\alpha_0 & \gamma_0
\end{pmatrix}
\begin{pmatrix}
1 &  0\\[6pt]
s & 1
\end{pmatrix}
\end{equation}
yielding
\begin{equation}
\beta(s)=\beta_0-2\alpha_0 s+\gamma_0 s^2
\end{equation}
and
\begin{equation}
\alpha(s)=\alpha_0-\gamma_0 s.
\end{equation}
We can conclude that the $\beta$-function in a drift follows a $s$-parabolic law.

\subsubsection{The differential relation between $\alpha$ and $\beta$}

Up to now, we recalled how to compute the $\alpha(s)$ and $\beta(s)$ at the start of the lattice ($\alpha(s_0)$ and $\beta(s_0)$) and how to propagate them along the lattice. We would like now to investigate if there is a differential relation between these two functions of the $s$-position.
We consider the general $\Delta M$ matrix for the infinitesimal quadrupole of length $\Delta s$,
\begin{equation}
\Delta M=\begin{pmatrix}
1 &        \Delta s\\[6pt]
-K(s)\Delta s & 1
\end{pmatrix}.
\end{equation}
Note that $\Delta M$ is just the product of a drift of length $\Delta s$ and a thin focusing quadrupole of gradient $K(s) \Delta s$ where we neglected the second order terms of $\Delta s$.  $\Delta M$ is then symplectic only for $\Delta s\rightarrow0$.
From Eq.~(\ref{eq:standardPropagation}) we have that 
\begin{equation}
\underbrace{
\begin{pmatrix}
\beta(s+\Delta s) & -\alpha(s+\Delta s)\\[6pt]
-\alpha(s+\Delta s)  & \gamma(s+\Delta s)
\end{pmatrix}}_{J(s+\Delta s)\Omega^{-1}}=
\Delta M\ 
\underbrace{
\begin{pmatrix}
\beta(s) & -\alpha(s)\\[6pt]
-\alpha(s)  & \gamma(s)
\end{pmatrix}}_{J(s)\Omega^{-1}}
\Delta M^T.
\label{eq:differentialTransport}
\end{equation}
Observing that
\begin{equation}
\lim_{\Delta s\rightarrow0} \dfrac{J(s+\Delta s)-J(s)}{\Delta s} \Omega^{-1}=
\begin{pmatrix}
\beta'(s) & -\alpha'(s)\\[6pt]
-\alpha'(s)  & \gamma'(s)
\end{pmatrix}
\label{eq:differentialJ}
\end{equation}
where we used standard notation $\dfrac{df}{ds}=f'$ and replacing Eq.~(\ref{eq:differentialTransport}) in Eq.~(\ref{eq:differentialJ}), one obtains
\begin{eqnarray}
\beta'(s)&=&-2 \alpha(s)\\\label{eq:derivative_of_beta}
\alpha'(s)&=&-\gamma+K(s) \beta(s).
\end{eqnarray}

Equation~(\ref{eq:derivative_of_beta}) represents the well-know relation between $\alpha$ and the derivative of $\beta$.
Replacing $\alpha$ and $\gamma$ in the latter equation with functions of $\beta$, it yields the \emph{non-linear differential equation}:
\begin{equation}
\boxed{\dfrac{\beta'' \beta}{2}-\dfrac{\beta'^ 2}{4}+K(s)\beta^2=1}.
\end{equation}

It is important to note that, even if we are discussing linear optics, the differential equation between $\beta$ and $K$ is strongly non-linear. Therefore, in order to avoid the linear tracking and to decompose the~problem in properties of lattice and properties of the beam we introduced new functions of the $s$-positions ($\alpha,~\beta,~\gamma$) that are related by a non-linear differential equation to the lattice gradients.


\subsubsection{EXAMPLE. From matrices to Hill's equation}
Following the notation already introduced
\begin{equation}
X(s+\Delta s)=\Delta M\ X(s)
\end{equation}
with $X(s)=(x(s), \dfrac{p_x(s)}{p_0})^T\underset{p_0\approx p_s}{\red{\approx}} (x(s),x'(s))^T$, therefore
\begin{equation}
X'(s)=
\begin{pmatrix}
x'(s)\\
x''(s)
\end{pmatrix}
=\lim_{\Delta s\rightarrow 0} \dfrac{X(s+\Delta s)-X(s)}{\Delta s}=
\begin{pmatrix}
x'(s)\\[6pt]
-K(s) x(s)
\end{pmatrix}
\end{equation}
one can find back the \red{Hill's equation} 
\begin{equation}
\boxed{x''(s)+K(s) x(s)=0}
\end{equation}
starting from a pure matrix approach, and without introducing directly the Lorentz force. This shows the~full equivalence of the two formalisms.

\subsection{Courant-Snyder  invariant}
 We showed in the previous sections, how to compute the optics functions of the lattice (that is, functions  independent on the particle initial conditions). In this Section we are going to investigate, given a particle with initial coordinate $X$, if and how we can define a $X$-dependent quantity that is conserved during the~motion of the particle in the machine. 
This invariant exists and is called \red{Courant-Snyder invariant} or \red{action} of the particle. It is defined as
\begin{equation}
\boxed{J_{CS}=\dfrac{1}{2}X^T \Omega\ J^{-1}\ X}.
\end{equation}
In fact from Eq.~(\ref{eq:standardPropagation}) we have that
\begin{equation}
\underbrace{\dfrac{1}{2}X_1^T \Omega\ J_1^{-1}\ X_1}_{J_{CS}(s_1)}= \dfrac{1}{2}X_0^T M^T (M\ J_0 \Omega^{-1}\ M^T)^{-1} M\ X_0=
\underbrace{\dfrac{1}{2}X_0^T \Omega\ J_0^{-1}\ X_0}_{J_{CS}(s_0)},
\end{equation}
i.e.,~$J_{CS}(s_1)=J_{CS}(s_0)$. It is worth noting that in \cite{CS} the invariant of motion is defined as $2\ J_{CS}$.
In~the~normalized phase-space, remembering that   $X=\bar{P}\ \bar{X}$, we have
\begin{equation}
\dfrac{1}{2}X^T \Omega\ J^{-1}\ X = \dfrac{1}{2}\bar{X}^T \underbrace{\bar{P}^T \Omega\ J^{-1} \bar{P}}_{I}\ \bar{X}=\dfrac{1}{2}\bar{X}^T\ \bar{X},
\end{equation}
that is the $J_{CS}$ is half of the square of the radius, $\rho$, defined by the particle initial position in the normalized phase-space. The \red{angle} of the particle is defined as the the particle initial angle, $\mu$, in the normalized phase-space and polar coordinates ($\rho$,$\mu$).
Hence, the normalized phase-space is also called \red{action-angle} space. 
From Listing~\ref{lst:CS}, going from the matrix form to the polynomial form, one finds back the definition of the invariant $J_{CS}$ as function of the optics functions

\begin{equation}
J_{CS}= \dfrac{\gamma}{2} x^2+\alpha\ x p_x + \dfrac{\beta}{2} p_x^2.
\label{eq:CSinvariant}
\end{equation}

It is worth noting that, under the assumptions of trace-space and phase-space equivalence (see Section~\ref{sec:referenceSystem}),  the invariant of motion can also be expressed in the trace-space variables as 
\begin{equation}
J_{CS}^{\rm trace-space}=\dfrac{\gamma}{2} x^2+\alpha\ x x' + \dfrac{\beta}{2} x'^2.
\label{eq:CSinvariantTS}
\end{equation}To be noted that Eq.~(\ref{eq:CSinvariantTS}) is not equivalent from a dimensional point of view to Eq.~(\ref{eq:CSinvariant}). Despite it, for the sake of simplicity and consistency with the existing conventions, we will use the same symbol $J_{CS}$ for both invariants.

\subsubsection{The betatron phase $\mu(s)$}
\label{sec:betatronPhase}
In normalized space, we just observed that the transport from $s$ to $s+\Delta s$ does not change  $J_{CS}$ but the~angle varies by $\Delta\mu=\mu(s+\Delta s)-\mu(s)$.

What is the $\Delta\mu$ introduced by a linear matrix $M=\begin{pmatrix}
m_{11} & m_{12} \\[6pt]
m_{21} & m_{22} 
\end{pmatrix}$?
To compute it we consider the~normalized phase-space
\begin{equation}
X(s)=\bar{P}(s)\ \bar{X}(s)\ \text{and}\ X(s+\Delta s)=\bar{P}(s+\Delta s)\ \bar{X}(s+\Delta s)
\end{equation}
and from
\begin{equation}
X(s+\Delta s)=M\ X(s),
\end{equation}
it yields
\begin{equation}
\bar{X}(s+\Delta s)=
\bar{P}(s+\Delta s)^{-1}\ M\ \bar{P}(s) \bar{X}(s)=\begin{pmatrix}
{\cos\Delta\mu} & {\sin\Delta\mu} \\[6pt]
-\sin\Delta\mu & \cos\Delta\mu
\end{pmatrix}\ \bar{X}(s).
\end{equation}
From the previous equation one gets
\begin{equation}
{\tan \Delta\mu}=\underbrace{\dfrac{{\sin \Delta\mu}}{{\cos \Delta\mu}}}_{\text{It does depend only on $\beta$ and $\alpha$ in $s$!}}{= \dfrac{m_{12}}{ m_{11}\ \beta(s)- m_{12}\ \alpha(s)}},
\label{eq:phase}
\end{equation}
that is the phase advance from $s$ to $s+\Delta s$.
The integer tune of the circular machine of length $C$ is defined as
\begin{equation}
\dfrac{1}{2\pi}~\mu(C),
\end{equation} therefore it represents the number of betatron oscillations between $s=0$ and $s=C$.
It is worth noting that between the $\mu_{OTM}$ (e.g., Eq.~(\ref{eq:rfactorization})) and $\mu(C)$ the following relation holds
\begin{equation}
\mu_{OTM}+2\pi k=\mu(C),
\end{equation}
where $k\in \mathbb{N}$ represents the integer number of betatron oscillations periods in the machine.

\subsubsection{EXAMPLE. The differential equation of  $\mu(s)$}
From the previous equation, if $M=
\begin{pmatrix}
1 & \Delta s \\[6pt]
- K(s) \Delta s & 1 
\end{pmatrix}$ ,  one gets
\begin{equation}
\mu' = \lim_{\Delta s\rightarrow 0} \dfrac{\tan\Delta\mu}{\Delta s}= \lim_{\Delta s\rightarrow 0}\dfrac{1}{\beta(s) -  \alpha(s)\ \Delta s}  = \dfrac{1}{\beta(s)},
\end{equation}
that is the well-known expression
\begin{equation}
\boxed{{\mu(s) =\int_{s_0}^{s} \dfrac{1}{\beta(\sigma)} {\rm d}\sigma + \mu(s_0)}}.
\end{equation}

\begin{figure}
    \centering
    \includegraphics[]{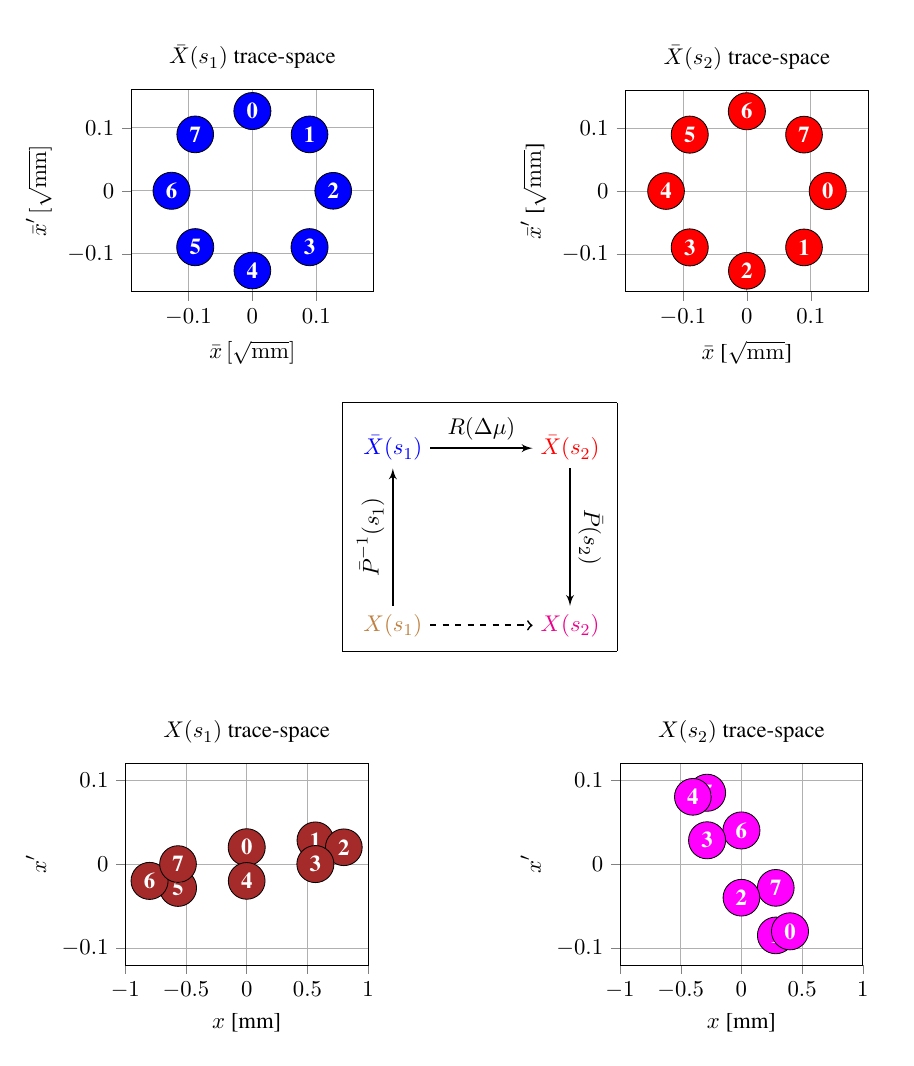}
\caption{Betatron oscillations from $s_1$ to $s_2$ in the physical and normalized trace-spaces (referred as $X$ and $\bar{X}$, respectively). In this example we assumed a tune of $\mu_{OTM}/2\pi=1/8$ and a phase advance between $s_1$ and $s_2$ of $\Delta\mu=\pi/2$. To be noted that all the transformations between these 4 trace-spaces are symplectic, therefore the $4\times8$ positions represented in the figure have the same $J_{CS}$. The number of the markers indicates the turn number.}
\label{fig:betatronOscillation}
\end{figure}

\subsubsection{EXAMPLE. The betatron oscillations}
\label{BetatronOscillation}
We can describe a betatron oscillation from $s_1$ to $s_2$ in terms of the Twiss parameters and the particle initial conditions.

This can be easily done by transforming the vector X in the normalized phase-space in $s_1$, moving it from $s_1$ to $s_2$ in the normalized space (pure rotation of the phase $\Delta\mu$) and back transform it in the~original phase-space, as shown in Fig.~\ref{fig:betatronOscillation}.


This is a very important result, since it implies that the motion of the particle is a pure rotation in the normalized phase-space also along the machine: this generalizes the result obtained from the R-factorization of the $M_{OTM}$.
As show in the Listing \ref{lst: mathematicaTransport}, one can express the $M$ matrix as function of the~optics function at $s_1$ and $s_2$, yielding 
\begin{eqnarray}
M&=& \bar P(s_2)\ R(\Delta\mu)\bar P(s_1)^{-1} = \\
&=&\begin{pmatrix}
\sqrt{\dfrac{\beta_2}{\beta_1}}(\cos \Delta\mu +\alpha_1 \sin \Delta\mu) & \sqrt{\beta_1 \beta_2} \sin\Delta\mu \\[12pt]
 \dfrac{\alpha_1-\alpha_2}{\sqrt{\beta_1 \beta_2}}\cos \Delta\mu -\dfrac{1+\alpha_1\alpha_2}{\sqrt{\beta_1\beta_2}} \sin \Delta\mu & 
\sqrt{\dfrac{\beta_1}{\beta_2}}(\cos\Delta\mu-\alpha_2 \sin \Delta\mu)
\end{pmatrix}.
\label{eq:transport}
\end{eqnarray}


\subsubsection{EXAMPLE. Solution of Hill's equation}
Remembering  the important result that motion of the particle is a pure rotation in the~normalized phase-space also along the machine, we can express the motion of the particle from its initial condition in the~normalized phase-space (i.e., action and initial phase that is $J_{CS}$ and $\mu_0$).  From the definition of $J_{CS}$ it follows that the radial position of the particle in the normalized phase-space is $\sqrt{2J_{CS}}$ and is angular position has a phase $\mu(s)$ in addition to the initial phase $\mu_0$. Remembering that, for positive $\mu$, the rotation is clockwise, one gets
\begin{eqnarray}
X(s)&=& \bar P(s) \begin{pmatrix}
\sqrt{J_{CS}} \cos(\mu(s)+\mu_0) \\[6pt]
-\sqrt{J_{CS}} \sin(\mu(s)+\mu_0) \\
\end{pmatrix}=\\
&=& \begin{pmatrix}
\sqrt{J_{CS}\beta(s)} \cos(\mu(s)+\mu_0) \\[6pt]
-\sqrt{\dfrac{J_{CS}}{\beta(s)}}[ \alpha(s) \cos(\mu(s)+\mu_0)+\sin(\mu(s)+\mu_0)]\\
\end{pmatrix}.
\end{eqnarray}
This is indeed the solution of the Hill's equation given the particle initial conditions $J_{CS}$ and $\mu_0$.

\subsection{EXAMPLE. From the CO matrix  to the CO formula.}
\label{COComputation}


Up to now we implicitly  assumed that the closed orbit (CO) corresponded to the reference orbit. This is not always true. In fact, during the machine operation one can switch on orbit correctors additional to the~ones defining the alignment of the magnetic elements.
Assuming a $M_{OTM}(s_0)$ and a single thin kick $\Theta$ at $s_0$ (independent from $X_m$) we can write
\begin{equation}
X_{m+1}(s_0)=M_{OTM}(s_0)\ X_{m}(s_0)\ {+}\ \Theta.
\label{eq:co}
\end{equation}
In the 1D case $\Theta$ can represent a kick of a dipole correction or misalignment of a quadrupole ($\Theta=(0, \theta)^T$). The closed orbit solution can be retrieved imposing $X_{m+1}=X_{m}$ (\red{fixed point} after 1-turn), yielding
\begin{equation}
{X_{n}(s_0)=(I-M_{OTM}(s_0))^{-1} \Theta(s_0)}.
\label{eq:CODefinition}
\end{equation}
From the Eq.~(\ref{eq:CODefinition}) we can find the fixed point in $s_0$. Please note that the CO derivative is discontinuous in $s_0$ so the previous formula refers to the CO after the kick. 
Solving the Eq.~(\ref{eq:CODefinition}) and transporting the~fixed point from $s_0$ to $s$ using Eq.~(\ref{eq:transport}) as shown in  Listing~\ref{lst: mathematicaCO} we found back the known equation of the~closed orbit
\begin{equation}
x_{CO}(s)=\dfrac{\sqrt{\beta(s)\beta(s_0)}}{2 \sin(\pi Q)}\theta_{s_0} \cos(\mu(s) -\pi Q),
\end{equation}
where $\mu(s)$ is the phase advance ($>0$) from $s_0$ to $s$. We can relax the last condition by replacing $\mu(s)$ with $|\mu(s)-\mu(s_0)|$. In presence of multiple $\theta(s_i)$ one can sum the different contributions along~$s$.

\subsection{Computing dispersion and chromaticity}
Up to now we considered all the optics parameters for the on-momentum particle. To evaluate the off-momentum effect of the closed orbit and the tune we introduce the dispersion, $D_{x,y}(s,  \dfrac{\Delta p}{p_0})$, and chromaticity, $\xi_{x,y}(\dfrac{\Delta p}{p_0})$, functions respectively, as
\begin{equation}
\Delta CO_{x, y}(s)={D_{x,y}}\left(s,  \dfrac{\Delta p}{p_0}\right)\times \dfrac{\Delta p}{p_0},\quad D_{x,y} (s+C)=D(s) 
\end{equation}
and 
\begin{equation}
\Delta Q_{x,y}={\xi_{x,y}} \left( \dfrac{\Delta p}{p_0}\right)\times\dfrac{\Delta p}{p_0}.
\end{equation}  

In order to compute numerically the $D_{x,y}$ and $\xi_{x,y}$  we can compute  the $CO_{x,y}$ and the $Q_{x,y}$ as function of $\dfrac{\Delta p}{p_0}$. To do that we have to compute $M_{OTM}(s,\dfrac{\Delta p}{p_0})$, that is evaluating the property of the element of the lattice as function of $\dfrac{\Delta p}{p_0}$.

\begin{itemize}
\item In a thin quadrupole the focal length linearly scales with the particle momentum:
\begin{equation}
\begin{pmatrix}
1 & 0\\[6pt]
-\dfrac{1}{{f\left(\dfrac{\Delta p}{p_0}\right)}} & 1
\end{pmatrix}{\rightarrow}
\begin{pmatrix}
1 & 0\\[6pt]
-\dfrac{1}{{f_0 \times \left(1+\dfrac{\Delta p}{p_0}\right)}} & 1
\end{pmatrix}.
\end{equation}
\item A dipolar corrector $\theta$, scales with the inverse of the beam rigidity:
\begin{equation}
\begin{pmatrix}
0\\
{\theta\left(\dfrac{\Delta p}{p_0}\right)}
\end{pmatrix}{\rightarrow}
\begin{pmatrix}
0\\[6pt]
{\dfrac{\theta_0}{1+\dfrac{\Delta p}{p_0}}} 
\end{pmatrix}.
\end{equation}
\item For the dipolar magnet defining the reference orbit (e.g., the arc dipole of a synchrotron) it is important to consider only the differential kick due to the off-momentum:
\begin{equation}
\begin{pmatrix}
0\\
{\theta\left(\dfrac{\Delta p}{p_0}\right)-\theta_0}
\end{pmatrix}{\rightarrow}
\begin{pmatrix}
0\\
{-\dfrac{\Delta p}{p_0}\dfrac{\theta_0}{1+\dfrac{\Delta p}{p_0}}} 
\end{pmatrix}.
\end{equation}

\end{itemize}

\section{Particle ensembles}

\subsection{The beam distribution}

\begin{figure}
\begin{center}
\input{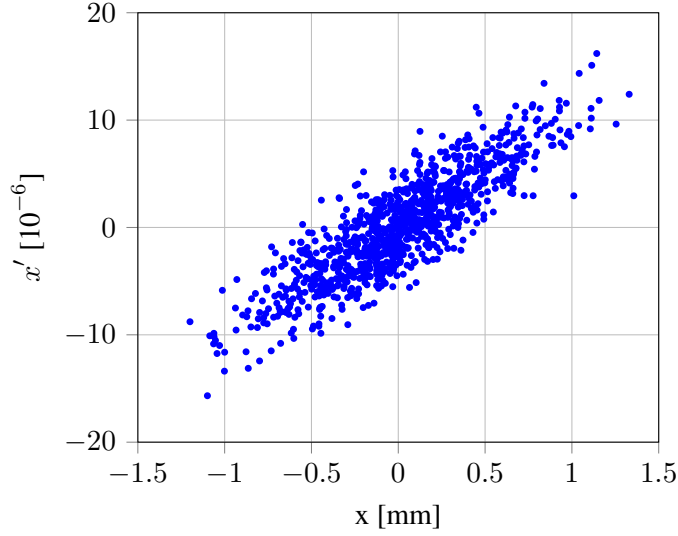}
\caption{The trace-space of an ensemble of particles.}
\label{eq:ensemble}
\end{center}
\end{figure}

The beam can be considered as a set of $N$ particles (Fig.~\ref{eq:ensemble}). 
To track $N$ particles we can use the same approach of the single particle tracking were $X$ becomes $X_{B}$, a $2n\times N$ matrix:
\begin{equation}
X_{B}=\begin{pmatrix}
X_1, X_2,\dots,X_n
\end{pmatrix}.
\end{equation}
We will restrict ourselves to the 1D case ($n=1$). We are looking for one or more statistical quantities that represents this ensemble and its evolution in the lattice.

A natural one is the average $J_{CS}$ over the ensemble:
\begin{equation}
\dfrac{1}{N} \sum_{i=1}^N J_{CS, i} =\langle J_{CS} \rangle.
\end{equation}
From the definition it follows that the quantity is preserved during the beam evolution along the linear lattice.

\subsubsection{The beam emittance}
We will see that  $\langle J_{CS} \rangle$ converges, under specific assumptions (see later), to the \blue{rms emittance} of the~beam, \blue{$\epsilon_{rms}$}
\begin{equation}
\boxed{\epsilon_{rms}=\sqrt{ \det(\underbrace{\dfrac{1}{N} X_{B} X_{B}^T}_{\text{\blue{$\sigma$  matrix}}}).}}
\label{eq:rmsEmittance}
\end{equation}
where $\dfrac{1}{N} X_{B} X_{B}^T$ represents the beam $\sigma$ matrix.

One can  see that the \red{$\epsilon_{rms}$ is preserved for the symplectic linear transformation $M$} from $s_0$ to $s_1$ (see Cauchy-Binet theorem):
\begin{eqnarray}
\epsilon_{rms}^2(s_0)&=& \det( \dfrac{1}{N} X_{B} X_{B}^T)\\
\epsilon_{rms}^2(s_1)&=& \det(  M\ \underbrace{\dfrac{1}{N} X_{B} X_{B}^T}_{\sigma(s_0)}\ M^T)=\underbrace{\det{M}}_{=1}\det(\dfrac{1}{N} X_{B} X_{B}^T) \underbrace{\det M^T}_{=1},
\end{eqnarray}
where $X_{B}$ denotes $X_{B}(s_0)$. Therefore, we have that 
\begin{equation}
\boxed{\sigma(s_1)=  M\ \sigma(s_0)\ M^T}
\label{eq:sigmaEvolution}
\end{equation}

and this transport equation is very similar to the one in Eq.~(\ref{eq:standardPropagation})
\begin{equation}
{{\begin{pmatrix}
\beta(s_1)  &        -\alpha(s_1) \\[6pt]
-\alpha(s_1)  & \gamma(s_1) 
\end{pmatrix}}}
 = M  
{{\begin{pmatrix}
\beta(s_0)  &        -\alpha(s_0)\\[6pt]
-\alpha(s_0) & \gamma(s_0)
\end{pmatrix}}}
\ M^T.
\label{eq:standardPropagation2}
\end{equation}

\subsubsection{The $\sigma$ matrix}
By the $\sigma$-matrix definition (Eq.~(\ref{eq:rmsEmittance})), it follows that (e.g., 2D trace-space)  
\begin{equation}
\boxed{
\sigma= \begin{pmatrix}
\dfrac{1}{N}\sum\limits_{i=1}^N x_i x_i & \dfrac{1}{N}\sum\limits_{i=1}^N x_i x'_i\\[15pt]
\dfrac{1}{N}\sum\limits_{i=1}^N x'_i x_i & \dfrac{1}{N}\sum\limits_{i=1}^N x'_i x'_i
\end{pmatrix}=
\begin{pmatrix}
\overbrace{\langle \bar{x}^2 \rangle}^{x_{rms}^2} & \langle x x' \rangle\\[6pt]
\langle x x'\rangle & \underbrace{\langle \bar{x}'^2 \rangle}_{x_{rms}^{'2}}\\
\end{pmatrix}}
\label{eq:statisticalMeaning}
\end{equation}
and therefore, we can write
\begin{equation}
\boxed{\epsilon_{rms}=\sqrt{\langle x^2 \rangle\langle x'^2 \rangle- \langle x x' \rangle^2}}. 
\end{equation}

To summarize, from the transport equation of the $\sigma$ matrix (Eq.~(\ref{eq:sigmaEvolution})) and from its statistical meaning (Eq.~(\ref{eq:statisticalMeaning})), we showed how numerically transport the second-order moments of the beam distribution. 

\subsection{Matched beam distribution}

A beam distribution is matched in $s_0$ to the specific optics functions ${\alpha(s_0)}$ and ${\beta(s_0)}$  if the corresponding normalized distribution $\bar{X}_B=\bar{P}^{-1} X_{B}$  is statistically invariant by rotation. In other words $\bar{X}_B$ has an~azimuthal symmetry therefore $\langle \bar{x} \bar{x}' \rangle=0$ and $\langle \bar{x}^2 \rangle=\langle \bar{x}'^2 \rangle$.  
An example of matched and mismatched beams are presented in Figs.~\ref{eq:matchedBeam} and \ref{eq:mismatchedBeam}, respectively.
\begin{figure}
\begin{center}
\input{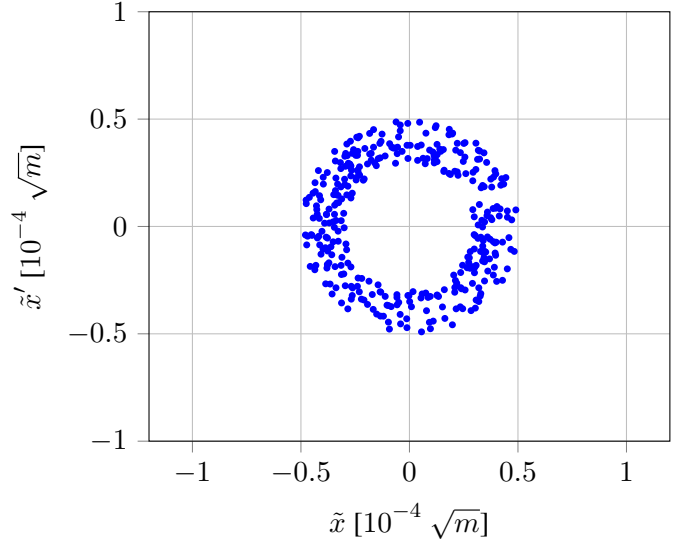}
\caption{A matched beam in the normalized phase-space.}
\label{eq:matchedBeam}
\end{center}
\end{figure}
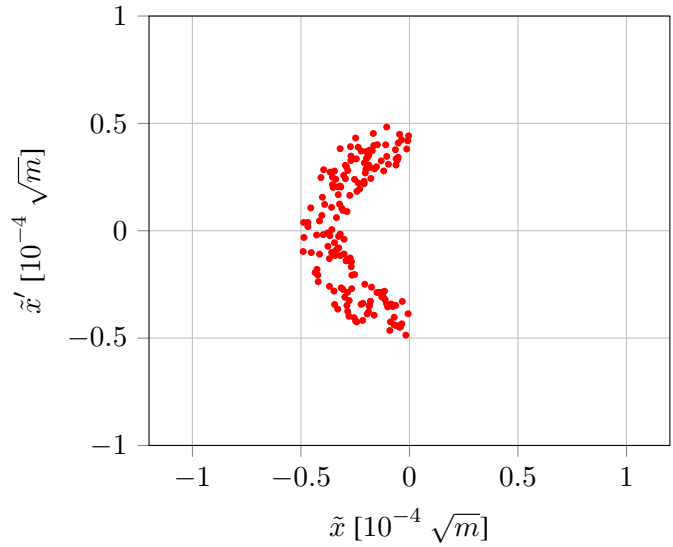
\begin{figure}
\begin{center}
\begin{tikzpicture}

\begin{axis}[
tick align=outside,
tick pos=left,
xlabel={$\tilde{x}$ [$10^{-4}$ $\sqrt{m}$]},
xmajorgrids,
xmin=-1.2, xmax=1.2,
ylabel={$\tilde{x}'$ [$10^{-4}$ $\sqrt{m}$]},
ymajorgrids,
ymin=-1., ymax=1.
]
\addplot [semithick, red, mark=*, mark size=1, mark options={solid}, only marks, forget plot]
table [row sep=\\]{%
-0.195070067491299	-0.387485692646493 \\
-0.306292754682705	0.0944135951571304 \\
-0.00614851454572258	-0.386653282462528 \\
-0.358440674089108	-0.10126889813277 \\
-0.0165965032727213	-0.486026105038529 \\
-0.297977527733133	-0.309056751153177 \\
-0.138529102384018	-0.287306057890213 \\
-0.151831035681602	0.297891878359912 \\
-0.295203703824072	0.242845065554368 \\
-0.203425401564785	0.269026819916578 \\
-0.0904310882702814	-0.464320469523672 \\
-0.162708189484917	-0.393448823257464 \\
-0.0412970175813061	-0.448127742066759 \\
-0.235486859635964	0.222365519326684 \\
-0.357899586138948	0.00546080968154961 \\
-0.311541398138992	0.10447876772088 \\
-0.127817856617517	-0.287454320735992 \\
-0.414576691721028	0.0457159233506373 \\
-0.0386237488409311	-0.437408838710159 \\
-0.199406692298626	0.368881873832684 \\
-0.325734152550001	-0.0801111704951369 \\
-0.287342811142315	0.0892677627877925 \\
-0.0840263490469311	-0.341365082362623 \\
-0.208585224446445	0.230253509022008 \\
-0.35138906824163	0.201892090464617 \\
-0.489801538987609	-0.0966444200753803 \\
-0.0472958201382572	-0.448915112526136 \\
-0.366436540554185	-0.0231093510750174 \\
-0.272088299310461	0.391705797426722 \\
-0.114919719344914	-0.28136528259912 \\
-0.291697291596765	-0.140226500760091 \\
-0.427843661894249	-0.020007064703499 \\
-0.316115927976924	0.203777107643416 \\
-0.339047578447157	-0.0915746964951337 \\
-0.0553500700186031	0.328733218942275 \\
-0.0361508571362214	0.423498818441892 \\
-0.369050993397157	-0.129589454570797 \\
-0.403514645179276	0.0713588616894318 \\
-0.274821674009225	0.164794151532245 \\
-0.105697949376366	-0.339932290291624 \\
-0.345575268944748	0.278070617341232 \\
-0.104860888393974	0.483073129316302 \\
-0.401079029007102	0.155912578961642 \\
-0.236254252842374	0.38894233662923 \\
-0.0873828728380465	-0.424633486642282 \\
-0.290120480198842	0.298170697024972 \\
-0.198028367861984	0.336709265041299 \\
-0.109375081009071	0.400257563311883 \\
-0.19055113580234	0.345745508576459 \\
-0.278840116291373	-0.398710963928846 \\
-0.30122425417025	-0.108294609946923 \\
-0.354794749571513	0.250232674048126 \\
-0.347621752721631	-0.281034195547046 \\
-0.330313298001948	-0.365126176358934 \\
-0.0127881728970626	0.380793299481914 \\
-0.0772623239712887	-0.35300787743541 \\
-0.205861658033271	-0.249266128596416 \\
-0.051856353375756	0.341878709136104 \\
-0.266337977927735	-0.269964131839474 \\
-0.164426197211984	0.396695204705509 \\
-0.187360671953859	0.30595659767016 \\
-0.422330246379888	-0.206194488510589 \\
-0.111991714652913	-0.321259232516511 \\
-0.2547625274959	-0.404582158427256 \\
-0.147875735805009	0.400643991580792 \\
-0.358844944184011	0.108950075322475 \\
-0.253012193721735	0.239408901708619 \\
-0.119100223707034	-0.311570754627582 \\
-0.396882298807922	-0.0181452638302386 \\
-0.161125281511921	0.288033316615829 \\
-0.320059177350057	0.207883442526472 \\
-0.345273640777938	-0.0557873517895314 \\
-0.100453198627962	-0.354715380118609 \\
-0.215360105031637	-0.417550874984243 \\
-0.0620672207381237	0.305989986903465 \\
-0.285512642356877	-0.286997226457913 \\
-0.0700705210364174	-0.40283863368405 \\
-0.303006292526328	0.257682712149073 \\
-0.273206426987853	-0.126797256390264 \\
-0.222148511948585	0.371409262649186 \\
-0.319222914653705	0.382151844678199 \\
-0.487612627115344	0.0387780442711999 \\
-0.183864026402041	-0.346754461123249 \\
-0.177656671116467	0.243413385379712 \\
-0.0353153407490433	-0.433474041379376 \\
-0.200758197550871	0.286841980469685 \\
-0.265211987923835	-0.207040544252775 \\
-0.0621063322535356	-0.441339665191561 \\
-0.0329276988641085	-0.329215737920595 \\
-0.322403299189515	0.12415109357602 \\
-0.45465566962142	0.10659985393442 \\
-0.17136602247861	0.372871273911179 \\
-0.182461784490773	0.295224125489663 \\
-0.00387775436471263	0.441899136693044 \\
-0.127562759085546	-0.307796732829393 \\
-0.214708177530268	-0.338037259620718 \\
-0.327966542233469	0.16780676516219 \\
-0.284095908305054	0.28042678013022 \\
-0.395460797689606	0.28440537987789 \\
-0.339792395593242	0.239866517883355 \\
-0.305992527971965	-0.273294013169957 \\
-0.125562363501647	-0.295172548024865 \\
-0.347178615879589	-0.103549650237042 \\
-0.0453455500230784	0.449382794124991 \\
-0.0706198662157636	-0.437944528690842 \\
-0.179617806446377	-0.328014752105127 \\
-0.330014634884807	0.204331971047956 \\
-0.096810599709761	0.309631232931464 \\
-0.43475713937533	-0.194540440288808 \\
-0.375666201521832	-0.072555558831876 \\
-0.242222917384192	0.183385617117444 \\
-0.0518667036889525	0.408336461294145 \\
-0.407680502970951	0.247546674963746 \\
-0.269428981341752	0.347633402420603 \\
-0.061481015751882	0.334590092763459 \\
-0.468173383773492	0.0386765942962225 \\
-0.185218257767498	-0.350941997195672 \\
-0.207707041734436	0.316235483283902 \\
-0.277647306878336	-0.324855226296249 \\
-0.287266761289272	-0.133016945593362 \\
-0.365646033107668	0.273373175025145 \\
-0.318991384352059	-0.0155679893116217 \\
-0.174807951225807	-0.262698810769063 \\
-0.149881234367519	-0.287596991863569 \\
-0.368580970253648	-0.258687455657989 \\
-0.288034767263077	-0.348375038728588 \\
-0.184693845299213	0.375115425690125 \\
-0.0648731665382942	0.377575008769118 \\
-0.344816508596993	-0.34339939516178 \\
-0.452333248871914	-0.101792022896052 \\
-0.00786338096157079	0.419060217866114 \\
-0.268214984883078	-0.167900483683602 \\
-0.301440856224057	-0.0394813558579351 \\
-0.229384959558228	0.194385771521758 \\
-0.222667920533628	-0.342740663002414 \\
-0.41416520248368	-0.109009529961613 \\
-0.270209801106923	0.326999303372557 \\
-0.329682664300712	-0.0819811769587342 \\
-0.318663817895597	-0.115683150285078 \\
-0.267191626673473	-0.144315860681779 \\
-0.314290408367173	-0.266110478858803 \\
-0.118813920394338	0.278254693004426 \\
-0.275614075873735	-0.140557808359451 \\
-0.242919733135453	-0.42428100413873 \\
-0.24805989712265	0.431835603023965 \\
-0.166691562244285	0.453221454970674 \\
-0.0634914307064189	-0.347587754558265 \\
-0.24823818755172	-0.419576814937518 \\
-0.343754464214002	-0.115478785150154 \\
-0.336261612816293	0.0611408321897584 \\
-0.420572238068229	-0.237200790885883 \\
-0.390763123420853	0.122727083519439 \\
-0.209085667516338	0.219077165983555 \\
-0.129973754735628	0.325498547659501 \\
-0.187448255663611	0.352377304865148 \\
-0.486164381338454	-0.03136026742579 \\
-0.104553877016801	0.346363581609818 \\
-0.284599814504834	-0.375432870261853 \\
-0.326442548203647	-0.0263903030953123 \\
-0.468046787127182	0.0187865175719138 \\
-0.244675405922848	0.334673731058909 \\
-0.252794606579664	-0.203720168176059 \\
-0.354548176096672	0.214410500800465 \\
-0.295102457881164	0.304883882166258 \\
-0.377167767880494	-0.00991589663340139 \\
-0.426026608368378	-0.179964073854194 \\
-0.188576571724051	-0.377666852610445 \\
};
\end{axis}

\end{tikzpicture}
\caption{A mismatched beam in the normalized phase-space.}
\label{eq:mismatchedBeam}
\end{center}
\end{figure}
It is worth noting that, since $\bar{P}^{-1}$ is a symplectic matrix,  $\bar{\epsilon}_{rms}=\epsilon_{rms}$ and, for a matched beam, we have
\begin{equation}
\bar{\sigma}=\dfrac{1}{N} \bar{X}_{B} \bar{X}_{B}^T = {\bar{P}^{-1} \sigma\ \bar{P}} =
\begin{pmatrix}
\overbrace{\langle \bar{x}^2 \rangle}^{\bar{x}_{rms}^2} & \overbrace{\langle \bar{x} \bar{x}' \rangle}^{=0}\\[6pt]
\underbrace{\langle \bar{x} \bar{x}' \rangle}_{=0} & \underbrace{\langle \bar{x}'^2 \rangle}_{\bar{x}_{rms}^{'2}}\\
\end{pmatrix}=
{\begin{pmatrix}
 \epsilon_{rms}   &  0 \\[6pt]
 0  &  \epsilon_{rms} \\
\end{pmatrix}}.
\end{equation} We can conclude that the normalized $\sigma$ matrix, $\bar{\sigma}$, is diagonal.

For a beam distribution  matched to the specific optics functions $\alpha(s_0)$ and $\beta(s_0)$ we have
\begin{equation}
\boxed{
\sigma= {\bar{P} 
\underbrace{\begin{pmatrix}
 \epsilon_{rms}   &  0 \\[6pt]
 0  &  \epsilon_{rms} \\
\end{pmatrix}}_{\bar{\sigma}}
\bar{P}^{-1}} =\epsilon_{rms}
\begin{pmatrix}
\beta(s_0)  & -\alpha(s_0) \\[6pt]
 -\alpha(s_0) & \gamma(s_0) 
\end{pmatrix}}
\label{eq:beamMatching},
\end{equation}
where we found back the rms beam size and divergence formulas, $\sqrt{\bar\beta \epsilon_{rms}}$ and $\sqrt{\bar\gamma \epsilon_{rms}}$, respectively. From Eqs.~(\ref{eq:standardPropagation2}) and (\ref{eq:beamMatching}), one can conclude that, if the beam is matched in the $s_0$-position, then is matched in all $s$. This implies that the second order statistical moments of a matched beam, in a periodic stable lattice and at given position $s$, are a turn-by-turn invariant.

Before concluding this chapter we demonstrate that, for matched beam, we have $\langle J \rangle=\epsilon_{rms}$. This is straightforward in the normalized phase-space, in fact from Eqs.~(\ref{eq:normalizedPS}) and (\ref{eq:CSinvariantTS})
\begin{equation}
J_{CS}=\dfrac{\bar{x}^2+\bar{x}'^2}{2}.   
\end{equation}
Since the beam is matched then $\langle \bar{x}^2 \rangle=\langle \bar{x}'^2 \rangle=\epsilon_{rms}$, yielding
\begin{equation}
\boxed{
\langle J_{CS}\rangle=\langle\dfrac{\bar{x}^2+\bar{x}'^2}{2}    \rangle=\dfrac{\langle \bar{x}^2 \rangle +\langle \bar{x}'^2 \rangle}{2}=\epsilon_{rms}}.
\end{equation}

\section{Conclusion}
In this Chapter we  recalled and summarized the main linear optics concepts of the accelerators beam dynamics theory with emphasis on the related computational aspects.
Using a pure linear algebra approach and via symplectic matrices transformations, we introduced the concepts of lattice stability, optics functions, normalized phase-space and invariant of motions. 
In addition to the dynamics of the single particle, we studied the ensembles of particles presenting the statistical invariant of the ensemble and the concept of beam matching. 

\appendix
\section{Code Listings}
For the convenience of the reader an electronic version of the Mathematica\texttrademark and Python3 listings can be found in \cite{mathematica} and \cite{python}, respectively.

\begin{lstlisting}[language=Mathematica, caption=The Mathematica\texttrademark ~input to compute the thick quadrupole matrix as limit of thin quadrupoles and drifts., label=lst:mathematicaLimit]
(* INPUT to Mathematica*)
MD{L_}={{1,L},{0,1}}
MQ{KL_}={{1,0},{-KL,1}}
FullSimplify[Limit[MatrixPower[MQ[ K1 L/n].MD[L/n], n], n -> \[Infinity], Assumptions -> { K1 > 0, L > 0}]]

(* OUTPUT *)
{{Cos[Sqrt[K1] L], Sin[Sqrt[K1] L]/Sqrt[
  K1]}, {-Sqrt[K1] Sin[Sqrt[K1] L], Cos[Sqrt[K1] L]}}
\end{lstlisting}

\begin{lstlisting}[language=Python, caption=Symbolic expression for $J$ matrix., label=lst: J ]
# INPUT to python3
from sympy import *
import numpy as np
m11=Symbol('m11'); m12=Symbol('m12'); m21=Symbol('m21'); m22=Symbol('m22')
omega=Matrix([[0,1], [-1,0]])
pbar=Matrix([[m11,m12], [m21,m22]])
J=pbar @ omega @ pbar.inv()
simplify(J.subs(m11*m22 - m12*m21,1))

# OUTPUT
Matrix([
[-m11*m21 - m12*m22,   m11**2 + m12**2],
[  -m21**2 - m22**2, m11*m21 + m12*m22]])
\end{lstlisting}

\begin{lstlisting}[language=Python, caption=Basic linear optics code., label=lst: basicLinearOptics]
# INPUT to python3
import numpy as np
la=np.linalg 
# Drift
def DRIFT(L=1):
    '''
    This a  matrix for a L-long drift. 
    '''
    return np.array([[1, L],[0, 1]])
# Quadrupole
def QUAD(f=1):
    '''
    This a matrix for a this quadrupole of focal lenght f.
    '''
    return np.array([[1, 0],[-1/f,1]])
# One turn maps of 100 m long FODO cell with 50 m focal length.
def M_OTM(f=50):
    M_OTM=DRIFT(50)@QUAD(-f)@DRIFT(50)@QUAD(f)
    return M_OTM

eigenvalues,P= la.eig(M_OTM())   # P is >>before the normalization<<
P=P/((la.det(P)*1j)**(1/len(P))) # P is >>after the normalization<<
D=np.diag(eigenvalues)
# Compute beta and alpha
beta=np.real(P[0,0]**2*2)
alpha=-np.real(P[1,0]*np.sqrt(2*beta))
print('The cell phase advance is ' + str(np.rad2deg(np.angle(D)[0][0])) + ' deg.')
print('The periodic beta at the start of the cell is '+ str(beta)+ ' m.')
print('The periodic alpha at the start of the cell is '+ str(alpha)+'.')

# OUTPUT
'The cell phase advance is 60.00000000000001 deg.'
'The periodic beta at the start of the cell is 173.20508075688772 m.'
'The periodic alpha at the start of the cell is -1.7320508075688774.'
\end{lstlisting}

\begin{lstlisting}[language=Python, caption=From the matrix to the polynominal form of the $J_{CS}$., label=lst:CS]
# INPUT to python3
from sympy import *
alpha=Symbol('alpha');beta=Symbol('beta');gamma=Symbol('gamma');
x=Symbol('x');px=Symbol('px')
omega=Matrix([[0,1], [-1,0]])
J=Matrix([[alpha, beta], [-gamma,-alpha]]); X=Matrix([[x],[px]])
expand(1/2*X.T@omega@J.inv()@X)[0,0].subs(alpha**2 - beta*gamma,-1)

# OUTPUT
1.0*alpha*px*x + 0.5*beta*px**2 + 0.5*gamma*x**2
\end{lstlisting}

\begin{lstlisting}[language=Python, caption=The phase advance computation., label=lst: phaseAdvance]
# INPUT to python3
import sympy as sy
beta0=sy.Symbol('beta0');alpha0=sy.Symbol('alpha0');
beta1=sy.Symbol('beta1');alpha1=sy.Symbol('alpha1');
m11=sy.Symbol('m11');m12=sy.Symbol('m12');m21=sy.Symbol('m21');m22=sy.Symbol('m22');

Pbar0=sy.Matrix([[sy.sqrt(beta0),0], [-alpha0/sy.sqrt(beta0),1/sy.sqrt(beta0)]])
Pbar1=sy.Matrix([[sy.sqrt(beta1),0], [-alpha1/sy.sqrt(beta1),1/sy.sqrt(beta1)]])
M=sy.Matrix([[m11, m12],[m21, m22]])
pprint(sy.simplify(Pbar1.inv()@M@Pbar0))

# OUTPUT
Matrix([[(-alpha0*m12 + beta0*m11)/(sqrt(beta0)*sqrt(beta1)), m12/(sqrt(beta0)*sqrt(beta1))], [(-alpha0*(alpha1*m12 + beta1*m22) + beta0*(alpha1*m11 + beta1*m21))/(sqrt(beta0)*sqrt(beta1)), (alpha1*m12 + beta1*m22)/(sqrt(beta0)*sqrt(beta1))]])
\end{lstlisting}

\begin{lstlisting}[language=Python, caption=Transport matrix as function of the optics parameter., label=lst: mathematicaTransport]
# INPUT to python3
import sympy as sy
beta1=sy.Symbol('beta1');alpha1=sy.Symbol('alpha1');
beta2=sy.Symbol('beta2');alpha2=sy.Symbol('alpha2');
phi=sy.Symbol('phi');Q=sy.Symbol('Q');theta1=sy.Symbol('theta1');
Pbar1=sy.Matrix([[sy.sqrt(beta1),0], [-alpha1/sy.sqrt(beta1),1/sy.sqrt(beta1)]])
Pbar2=sy.Matrix([[sy.sqrt(beta2),0], [-alpha2/sy.sqrt(beta2),1/sy.sqrt(beta2)]])
R=sy.Matrix([[sy.cos(phi),sy.sin(phi)], [-sy.sin(phi),sy.cos(phi)]])
sy.simplify(Pbar2@R@Pbar1.inv())

# OUTPUT
Matrix([[sqrt(beta2)*(alpha1*sin(phi) + cos(phi))/sqrt(beta1), sqrt(beta1)*sqrt(beta2)*sin(phi)], [(-alpha1*alpha2*sin(phi) + alpha1*cos(phi) - alpha2*cos(phi) - sin(phi))/(sqrt(beta1)*sqrt(beta2)), sqrt(beta1)*(-alpha2*sin(phi) + cos(phi))/sqrt(beta2)]])
\end{lstlisting}

\begin{lstlisting}[language=Python, caption=Closed orbit computation., label=lst: mathematicaCO]
# INPUT to python3
import sympy as sy
beta1=sy.Symbol('beta1');alpha1=sy.Symbol('alpha1');
beta2=sy.Symbol('beta2');alpha2=sy.Symbol('alpha2');
Q=sy.Symbol('Q');theta1=sy.Symbol('theta1');phi=sy.Symbol('phi')

J=sy.Matrix([[alpha1, beta1],[-(1+alpha1**2)/beta1,-alpha1]])
I=sy.Matrix([[1, 0],[0,1]])
MCO=sy.simplify((I-(I*sy.cos(2*sy.pi*Q)+J*sy.sin(2*sy.pi*Q))).inv())
X0=sy.simplify(MCO@sy.Matrix([[0],[theta1]]))
T=sy.Matrix([[(sy.sqrt(beta2)*(sy.cos(phi)+alpha1*sy.sin(phi)))/sy.sqrt(beta1),\
              sy.sqrt(beta1)*sy.sqrt(beta2)*sy.sin(phi)],\
            [-((-alpha1+alpha2)*sy.cos(phi)+sy.sin(phi)+alpha1*alpha2*sy.sin(phi))/sy.sqrt(beta1)/sy.sqrt(beta2),\
            sy.sqrt(beta1)*(sy.cos(phi)-alpha2*sy.sin(phi))/sy.sqrt(beta2)]])
sy.simplify(T@X0)

# OUTPUT
Matrix([[sqrt(beta1)*sqrt(beta2)*theta1*(sin(phi) + cos(phi)/tan(pi*Q))/2], [-sqrt(beta1)*theta1*(alpha2*sin(phi) + alpha2*cos(phi)/tan(pi*Q) + sin(phi)/tan(pi*Q) - cos(phi))/(2*sqrt(beta2))]])
\end{lstlisting}

\end{document}